\documentclass[preprint, 5p, 12pt, twocolumn]{elsarticle}
\usepackage{amsthm}

\makeatletter
\def\els@aparagraph[#1]#2{\elsparagraph[#1]{#2\@addpunct{}}}
\def\els@bparagraph#1{\elsparagraph*{#1\@addpunct{}}}
\makeatother

\usepackage{graphicx}
\usepackage{xcolor}
\usepackage{xspace}
\usepackage{listings}
\usepackage{multirow}
\usepackage[frozencache=true,cachedir=mintedcache]{minted}
\usepackage{subcaption}
\usepackage[linesnumbered,plain,figure]{algorithm2e}

\usepackage{acronym}

\acrodef{AST}{Abstract Syntax Tree}
\acrodef{AOP}{Aspect-Oriented Programming}

\newcommand{\revblue}[1]{{\color{black}#1}\xspace}
\newcommand{\revred}[1]{{\color{black}#1}\xspace}

\newcommand{\sch}{SC\xspace}
\newcommand{\tra}{TR\xspace}
\newcommand{\jdk}{Java JDK\xspace}

\journal{Journal of Systems and Software}

\usepackage{hyperref}
\usepackage{cleveref}

\begin{document}
\begin{frontmatter}

\title{METFORD -- Mutation tEsTing Framework fOR anDroid}

% AUTHORS
\author[addrA,addrC]{Auri M. R. Vincenzi}
\ead{auri@ufscar.br,auri@fe.up.pt}

\author[addrA,addrC]{Pedro H. Kuroishi}
\ead{phk@ufscar.br}

\author[addrB]{João C. V. M. Bispo}
\ead{jbispo@fe.up.pt}

\author[addrC]{Ana R. C. da Veiga}
\ead{up202103388@edu.fe.up.pt}

\author[addrC]{David R. C. da Mata}
\ead{up202003557@edu.fe.up.pt}

\author[addrC]{Francisco B. Azevedo}
\ead{up201809186@edu.fe.up.pt}

\author[addrB]{Ana C. R. Paiva}
\ead{apaiva@fe.up.pt}

\address[addrA]{Department of Computing, Federal University of São Carlos, São Carlos -- SP -- Brazil}
\address[addrB]{INESC TEC, Faculty of Engineering, University of Porto, Porto, Portugal}
\address[addrC]{Faculty of Engineering, University of Porto, Porto, Portugal}

\begin{abstract}
Mutation testing may be used to guide test case generation and as a technique to assess the quality of test suites. Despite being used frequently, mutation testing is not so commonly applied in the mobile world. One critical challenge in mutation testing is dealing with its computational cost. Generating mutants, running test cases over each mutant, and analyzing the results may require significant time and resources. This research aims to contribute to reducing Android mutation testing costs. It implements mutation testing operators (traditional and Android-specific) according to mutant schemata (implementing multiple mutants into a single code file). It also describes an Android mutation testing framework developed to execute test cases and determine mutation scores. Additional mutation operators can be implemented in JavaScript and easily integrated into the framework. The overall approach is validated through case studies showing that mutant schemata have advantages over the traditional mutation strategy (one file per mutant). The results show mutant schemata overcome traditional mutation in all evaluated aspects with no additional cost: it takes 8.50\% less time for mutant generation, requires 99.78\% less disk space, and runs, on average, 6.45\% faster than traditional mutation. Moreover, considering sustainability metrics, mutant schemata have 8,18\% less carbon footprint than traditional strategy.
\end{abstract}

\end{frontmatter}

\section{Introduction}\label{sec:introduction}

We are increasingly dependent on mobile applications in our daily lives, and the number of mobile applications at our disposal is constantly growing. According to \citet{Buildfire23}, there are 4.83 million apps available for download (App Store has 1.96 million, and Google Play Store has 2.87). Therefore, it is of paramount importance to ensure the quality of these applications. However, it is known that mobile apps have several deficiencies, and several studies reveal them~\cite{9153947}.

One way to contribute to improving software quality is through software testing. According to \citet{10.1145/3526099} ``Mutation-based testing is a testing approach that uses mutants as test goals to create or improve a test suite.''. It is a fault injection white-box testing technique where minor modifications are inserted on purpose into the program by applying syntactic rules (mutation operators) to generate mutants (slightly different versions of the original program that mimic common programming faults). The final goal is to improve or create a test suite that can kill those mutants, i.e., obtain a different result when executing the original and modified versions of the program under test.

It is known that Android testing is different from traditional Java testing~\cite{8009912} because Android apps have specific characteristics, such as activities and intents. But, according to \citet{8009912}, ``Android mutation testing is significantly effective at detecting faults''.

Mutation-based testing is a computationally expensive technique. To be effectively applied, we must overcome several challenges spread through the different stages of the mutation-based testing process: generation of mutants, generation and execution of test cases, and analysis of the results~\cite{Bokaei19,Wong01MTNC}.
There are some strategies to reduce the cost of mutation-based testing. At a high level, cost reduction techniques can be classified into three distinct approaches: do less, do smarter, or do faster~\cite{Offutt2001}.

The overall cost of mutation-based testing on Android can be very high~\cite{DBLP:conf/seke/DengO18}, raising even more challenges than testing traditional software and compromising its practical use. One of those challenges is the time required to execute test cases. In a systematic mapping study, \citet{SilvaSTVR22} identified eight tools supporting mutation testing for Android applications: BacterioWeb~\cite{7899048}, DroidMutator~\cite{10.1145/3377812.3382134}, Edroid~\cite{8645883}, MDroid+~\cite{10.1145/3183440.3183492}, $\mu$Droid~\cite{10.1145/3106237.3106244}, MUTAPK~\cite{DBLP:conf/kbse/Escobar-Velasquez19, EscobarVelasquez20MAPK2ATRMTAA}, and two versions of muDroid~\cite{Deng15TMAAA, Wei15MDMTAA}. One of the main reasons for the high cost is the need to send each mutant to be executed in the device or emulator individually.

The METFORD framework aims to reduce mutation-based testing costs by implementing a set of mutation operators considering the Mutant Schemata strategy, encoding all mutants in one parameterized program or application (app). The authors believe that using meta mutants will reduce execution time and disk space compared with traditional mutants.

Mutant Schemata is a strategy designed to reduce the total cost of mutation testing. The central idea of this technique is to compose different programs into a metaprogram (all program versions are included in a single file). A control mechanism must be implemented to determine which program versions included in the schema are to be executed. The pioneering work regarding Mutant Schema is from \citet{Untch1993MutationAU}, who created a mutant schema generator for C. They used \textit{metamutants} and \textit{metaprocedures}. A \textit{metamutant} contains all the mutants in a single file. Our tool implements both strategies, Mutant Schemata (\sch) and Traditional Mutation (\tra), making it possible to compare both.

This paper presents an experimental study that assesses whether we can improve mutation-based testing efficiency by applying mutation testing to mobile applications. The experiment implements all mutants in one file (meta mutant), hoping to reduce memory costs and the time required for mutant generation and test case execution. We investigate alternatives to implement metamutants using a different strategy than the one described by~\cite{https://doi.org/10.1002/stvr.1769}, supported by the BacterioWeb mutation tool~\cite{7899048}, to improve performance.

We answer two research questions: RQ1: Do mutant schemata and traditional mutation strategies implementations exhibit the same behavior?; 
and RQ2: How do time and memory consumption compare in the two implementations (mutant schemata and traditional)?

The study's results reveal that a set of 14 mutation operators was successfully implemented using the mutant schemata strategy, and they can generate meta mutants with the same behavior as traditional mutants. Notably, on average, the mutant schemata strategy demands 8.50\% less time for mutant generation and 99.78\% less disk space. Regarding cost reduction related to execution time, mutant schemata surpass traditional mutation strategy by over 6.45\%, corresponding to the time for individual mutants to be copied, compiled, and loaded in mobile individually in the case of traditional mutation.

In summary, this paper makes the following contribution:

\begin{itemize}
    \item The METFORD framework, which supports mutation testing on Android mobile applications considering mutant schemata and traditional mutation strategies;

    \item An experimental study with six real Android mobile applications comparing both strategies and pointing out their advantages and disadvantages;

    \item Statistical data on Android-specific mutation operators for future framework extensions; 
    
    \item Make the tool, scripts, data, and mobile applications available in a repository\footnote{\url{https://github.com/specs-feup/paper-metford}} so other researchers can use and evolve it.
\end{itemize}

This paper is organized as follows: Section~\ref{sec:related-work} presents related works; Section~\ref{sec:metaford} describes our framework (METFORD) and describes how we implement meta mutants; Section~\ref{sec:case-study} presents the experiment description, collected data and answer for our research questions; Section~\ref{sec:threats} discusses the threats of validity. Finally, Section~\ref{sec:conclusions} presents the conclusions.

\section{Related Work}\label{sec:related-work}

\citet{Pizzoleto19ASLRTMRCMT} presented a systematic literature review to assess and categorize 157 studies focusing on the cost reduction of mutation testing. The studies were classified according to six main goals for cost reduction (see Table~\ref{tab:goalsCostReduction}) and 21 techniques (e.g., random mutation, higher order mutation, and mutant schemata). Moreover, the authors proposed 18 cost-reduction metrics to evaluate the studies regarding cost savings and mutation scores. 

\begin{table}[!htb]
\caption{Six main goals for cost reduction of mutation testing according to~\citet{Pizzoleto19ASLRTMRCMT}.}
\label{tab:goalsCostReduction}
\resizebox{\columnwidth}{!}{%
\begin{tabular}{|l|l|}
\hline
\textbf{ID} & \textbf{Goals for Cost Reduction} \\ \hline
1 & Reducing the number of mutants \\ \hline
2 & Automatically detecting equivalent mutants \\ \hline
3 & Executing Faster \\ \hline
4 & Reducing the number of test cases or the number of executions \\ \hline
5 & Avoiding the creation of certain mutants \\ \hline
6 & Automatically generating test casers \\ \hline
\end{tabular}%
}
\end{table}

The technique in this paper is an example of an Android mutant schemata encoding all mutants in one program to improve the execution time and reduce required storage. Following a mutant schemata strategy has additional benefits for Android testing since mutants are implemented in one file, which requires deploying the mutated app once to a device for running the tests.

Mutation testing for mobile applications has started to gain attention in recent years~\cite{SilvaSTVR22}. Given the specific characteristics of mobile devices and applications, mutation operators must be designed to model particular defects related to the mobile ecosystem~\cite{LinarezVasquez17EMTAA}. The systematic mapping study of~\citet{SilvaSTVR22} assessed 16 primary studies focusing on mutation for mobile applications and found 138 existing mutation operators categorized as configuration, connectivity, GUI, intent, location, persistence, sensors, and traditional operators.~\citet{LinarezVasquez17EMTAA} proposed a taxonomy of the most common Android bugs and categorized them as activities and intents, back-end services, collections and strings, data/objects parsing and format, threading, Android programming, non-functional requirements, GUI, I/O, device/emulator, API and libraries, connectivity, database and general programming. Moreover, the authors defined 38 mutation operators based on the proposed taxonomy.

We found few studies investigating cost reduction for Android mutation testing.~\citet{DBLP:conf/seke/DengO18} proposed a cost-reduction approach for Android mutation testing by identifying redundancy among mutation operators, i.e., mutation operators that fail to contribute to the quality of the tests. The authors evaluated 32 mutation operators (17 Android-specific and 15 Java-specific operators) and found that three Java-specific operators are redundant and can be discarded. Moreover, four Android-specific operators may also be excluded due to the operators' subsumption relation~\cite{Kurtz16AVSM}. \citet{https://doi.org/10.1002/stvr.1769} designed a mathematical model to estimate the time required for mutation testing on a mobile app by combining Mutant Schemata, Parallel Execution (i.e., executing mutation testing on multiple devices), and Only Alive (i.e., run tests on the remaining live mutants) to validate their proposal of assessing the improvement of testing time. 

Regarding mutation testing tools, the systematic mapping study of~\citet{SilvaSTVR22} identified eight tools supporting mutation testing for Android applications: BacterioWeb~\cite{7899048}, DroidMutator~\cite{10.1145/3377812.3382134}, Edroid~\cite{8645883}, MDroid+~\cite{10.1145/3183440.3183492}, $\mu$Droid~\cite{10.1145/3106237.3106244}, MUTAPK~\cite{DBLP:conf/kbse/Escobar-Velasquez19, EscobarVelasquez20MAPK2ATRMTAA}, and two versions of muDroid~\cite{Deng15TMAAA, Wei15MDMTAA}. Mutation testing may be applied at the source code level or APK level, which consists of decompiling the application under test to an Android assembly format (e.g., SMALI), mutating the assembly code, and recompiling the mutated version~\cite{DBLP:conf/kbse/Escobar-Velasquez19, EscobarVelasquez20MAPK2ATRMTAA}. It is worth noting that most tools only support mutant generation while few support mutant execution and analysis~\cite{SilvaSTVR22}. 

As far as we know, BacterioWeb is the only one implementing the mutant schemata strategy, but as per~\cite{SilvaSTVR22}, the tool is not publicly available.

\section{METFORD}\label{sec:metaford}

METFORD stands for Mutation tEsTing Framework fOR anDroid. We intend to provide a mutation testing tool that allows mutation testing on Android mobile applications and can be used not only as a proof of concept but also for experimentation, technology transference, and teaching.

\subsection{Mutation Strategies}

METFORD implements two mutation strategies, traditional and schemata. Each strategy is implemented as a separate script, and both strategies follow the same general algorithm:

\begin{itemize}
    \item Iterate over each source file;
    \item For each source file, iterate over all \ac{AST} nodes;
    \item For each \ac{AST} node, iterate over all selected mutation operators;
    \item Add the node to the operator;
    \item While there are mutations in the operator, mutate the code, store the necessary information, and restore the mutated code.
\end{itemize}

The main difference is that the traditional mutation strategy immediately generates a version of the project with the mutated code after each mutation. In contrast, the mutant schemata strategy generates a single project that contains all the mutations, and we control the mutation activation by a parameter.

Additionally, the mutant schemata strategy might not directly support certain specific mutations. As an example, the schemata strategy might not be able to apply a binary operator mutation directly over the code \texttt{int a = 1 + 2;} since the modifications could hide the declaration of the variable \texttt{a} to the code after the mutation, as illustrated in \Cref{fig:schemata_decl}, considering we are applying an arithmetic operator which replaces `\texttt{+}' per `\texttt{-}.'

\begin{figure}[!htb]
    \centering
\begin{minted}[frame=lines,fontsize=\scriptsize,samepage=true]{java}
switch(MUID_STATIC) {
  case 1: {
    int a = 1 - 2;
    break;
  }
  default: {
    int a = 1 + 2;
}
System.out.println("a: " + a);
\end{minted}
 \caption{Example of how applying directly the schemata strategy can introduce compilation errors.}
    \label{fig:schemata_decl}
\end{figure}

There are several ways to deal with situations like this: (1) specialized mutation operations aware of the schemata strategy; (2) transform the code to enable the application of the mutation operator.

We consider it out of scope for the present work to explore in what cases the mutant schemata strategy can be applied versus the traditional mutation strategy and what techniques can bring the schemata closer to the traditional mutation strategy.
For this reason, we only considered mutations that could be applied in both strategies.

There was, however, a transformation that we decided to apply to the code to enable more mutant schemata strategy implementations, as illustrated in Figure~\ref{fig:schemata_decl_transformation}. This was to decompose variables initialization, for example, transform the previous code to \texttt{int a; a = 1 + 2;}.

\begin{figure}[!htb]
    \centering
\begin{minted}[frame=lines,fontsize=\scriptsize,samepage=true]{java}
int a;
switch(MUID_STATIC) {
  case 1: {
    a = 1 - 2;
    break;
  }
  default: {
    a = 1 + 2;
}
System.out.println("a: " + a);
\end{minted}
 \caption{Example of declaration transformation to support mutant schemata strategy.}
    \label{fig:schemata_decl_transformation}
\end{figure}

\subsection{Mutant Schemata Implementation}

While the concept of mutant schemata is relatively straightforward (i.e., introducing a control-flow structure that allows the user to choose which mutation to execute), there are several ways to implement it, with different performance trade-offs.
Since the schemata strategy can potentially introduce a significant amount of new code and control flow, our concern relates to the impact on the application's execution time when running the tests.

For instance, BacterioWeb~\cite{7899048} uses static methods for implementing mutant schemata, which, besides reducing the code inserted in the original program, may increase the running time due to context saving of calling a function to activate each mutation. We adopted a different approach in our implementation, eliminating the need for additional methods calls.

Considering a global property to control the mutation to execute, we did a comparative analysis of different implementations of the schemata strategy along three main dimensions:  
(1) the property is stored as a final (e.g., \mintinline{java}|final static int MUID|) or non-final static variable (e.g., \mintinline{java}|static int MUID|), (2) the property is encoded as a \texttt{String} (e.g., \mintinline{java}|final static String MUID|) or an \texttt{int} (e.g., \mintinline{java}|final static int MUID|), and (3) the control structure is a \texttt{switch} (e.g., \mintinline{java}|switch(MUID) //...|) or an  \texttt{if-else} chain (e.g., \mintinline{java}|if(MUID == 1) //...|). 

Tables~\ref{tab:benchmark-jdk11} to~\ref{tab:benchmark-jdk21} illustrate the results of a micro-benchmark we carried out to decide the best implementation option of schemata strategy along these dimensions. The micro-benchmark simulates a scenario where 100 mutations are applied to the same statement using the schemata strategy, originating 101 possible paths (100 mutations plus the no-mutation path).

Instead of showing the results for all possible combinations of the three dimensions, we present the results for the combination that performed better (\texttt{static}, \texttt{int}, \texttt{switch}) and the results when one of the dimensions changes while keeping the values of the other dimension the same as the best combination.

Additionally, for each combination, we show two variants when the property selects either the first branch (the first \texttt{if} path or the first \texttt{switch case}) or the last branch (the last \texttt{else} or the \texttt{switch} default).

In all cases, the property is set by using \texttt{adb} \texttt{setprop(debug.MUID)}\footnote{\texttt{debug.MUID} is the name of the variable we use to set the mutant number to be executed.}. Using \texttt{debug.MUID} property makes testing easier since it does not require rooted devices for testing debug or release versions of the app.

Table~\ref{tab:benchmark-jdk11},  Table~\ref{tab:benchmark-jdk17}, and Table~\ref{tab:benchmark-jdk21} have the same structure. The benchmark for each implementation option is run 30 times. Column SUM contains the total time of these 30 runs in milliseconds. The MAX and MIN columns have the maximum and minimum running time among those 30 executions, and AVG is the average running time. Column INC corresponds to the time increment of the current strategy concerning our baseline implementation, comparing the first and last branches of both alternatives.

For a given modifier (first column) and property data type (second column), we have the time required to decide which mutant to execute when it matches the first or the last branch in a given control flow structure (third column). For instance, in Table~\ref{tab:benchmark-jdk11}, it is possible to see in the first row that we need 1874ms (in 30 executions) to select an operator when it matches the first switch case branch. The maximum time required was 125ms, the minimum was 36ms, and, on average, 62.5ms. This results in an increment (INC column) in time of 72.1\% compared to the baseline we defined after the data collection. Baseline uses a \texttt{static final} modifier, property variable of \texttt{int} type, and is implemented with a \texttt{switch}. The baseline first switch case branch took 36.3ms to be solved, on average.

\begin{table}[!htb]
    \caption{Benchmark for evaluating schemata implementation strategy - \jdk 11.}\label{tab:benchmark-jdk11}
    \centering
\resizebox{\columnwidth}{!}{%    
    \begin{tabular}{|l|l|l|r|r|r|r|r|}\hline
\multicolumn{3}{|c|}{\bf Schemata Alternatives} &
        \multirow{2}{*}{\bf SUM} &
	  \multirow{2}{*}{\bf MAX} &
	\multirow{2}{*}{\bf MIN} &
	\multirow{2}{*}{\bf AVG} & 
        \multirow{2}{*}{\bf INC (\%)}\\\cline{1-3}
\bf Modifier &
	\bf Prop. &
	\bf Branch &
	&
	&
	&
        &
        \\\hline
\multirow{2}{*}{static non-final} &
	 \multirow{2}{*}{int} &
	 first switch case &
	1874 &
	125 &
	36 &
	62.5 &
	72.1\\\cline{3-8}
 &
	  &
	 switch default &
	1399 &
	81 &
	43 &
	46.6 &
	22.4\\\hline
\multirow{2}{*}{static final} &
	 \multirow{2}{*}{int} &
	 first if path &
	1037 &
	43 &
	32 &
	34.6 &
	-4.8\\\cline{3-8}
 &
	  &
	 last else path &
	8736 &
	310 &
	280 &
	291.2 &
	664.3\\\hline
\multirow{2}{*}{static final} &
	 \multirow{2}{*}{String} &
	 first switch case &
	8878 &
	348 &
	285 &
	295.9 &
	715.2\\\cline{3-8}
 &
	  &
	 switch default &
	9503 &
	513 &
	306 &
	316.8 &
	731.4\\\hline
\multicolumn{8}{|c|}{\bf Baseline}\\\hline
\bf Modifier &
	\bf Prop. &
	\bf Branch &
        \bf SUM &
	  \bf MAX &
	\bf MIN &
	\bf AVG & 
        \bf INC (\%)\\\hline
\multirow{2}{*}{static final} &
	 \multirow{2}{*}{int} &
	 first switch case &
	1089 &
	43 &
	35 &
	36.3 &
	0.0\\\cline{3-8}
 &
	  &
	 switch default &
	1143 &
	42 &
	37 &
	38.1 &
	0.0\\\hline
\end{tabular}
}
\end{table}

\begin{table}[!htb]
    \caption{Benchmark for evaluating schemata implementation strategy - \jdk 17.}\label{tab:benchmark-jdk17}
    \centering
\resizebox{\columnwidth}{!}{%    
    \begin{tabular}{|l|l|l|r|r|r|r|r|}\hline
\multicolumn{3}{|c|}{\bf Schemata Alternatives} &
        \multirow{2}{*}{\bf SUM} &
	  \multirow{2}{*}{\bf MAX} &
	\multirow{2}{*}{\bf MIN} &
	\multirow{2}{*}{\bf AVG} & 
        \multirow{2}{*}{\bf INC (\%)}\\\cline{1-3}
\bf Modifier &
	\bf Prop. &
	\bf Branch &
	&
	&
	&
        &
        \\\hline
\multirow{2}{*}{static non-final} &
	 \multirow{2}{*}{int} &
	 first switch case &
	2042 &
	97 &
	45 &
	68.1 &
	61.2\\\cline{3-8}
 &
	  &
	 switch default &
	1265 &
	48 &
	40 &
	42.2 &
	-11.4\\\hline
\multirow{2}{*}{static final} &
	 \multirow{2}{*}{int} &
	 first if path &
	1113 &
	41 &
	34 &
	37.1 &
	-12.1\\\cline{3-8}
 &
	  &
	 last else path &
	8757 &
	306 &
	281 &
	291.9 &
	513.2\\\hline
\multirow{2}{*}{static final} &
	 \multirow{2}{*}{String} &
	 first switch case &
	8373 &
	289 &
	277 &
	279.1 &
	560.9\\\cline{3-8}
 &
	  &
	switch default &
	9545 &
	458 &
	309 &
	318.2 &
	568.4\\\hline
\multicolumn{8}{|c|}{\bf Baseline}\\\hline
\bf Modifier &
	\bf Prop. &
	\bf Branch &
        \bf SUM &
	  \bf MAX &
	\bf MIN &
	\bf AVG & 
        \bf INC (\%)\\\hline 
\multirow{2}{*}{static final} &
	 \multirow{2}{*}{int} &
	 first switch case&
	1267 &
	122 &
	39 &
	42.2 &
	0.0\\\cline{3-8}
 &
	  &
	 switch default &
	1428 &
	154 &
	42 &
	47.6 &
	0.0\\\hline

\end{tabular}
}
\end{table}

\begin{table}[!htb]
    \caption{Benchmark for evaluating schemata implementation strategy - \jdk 21.}\label{tab:benchmark-jdk21}
    \centering
\resizebox{\columnwidth}{!}{%    
    \begin{tabular}{|l|l|l|r|r|r|r|r|}\hline
\multicolumn{3}{|c|}{\bf Schemata Alternatives} &
        \multirow{2}{*}{\bf SUM} &
	  \multirow{2}{*}{\bf MAX} &
	\multirow{2}{*}{\bf MIN} &
	\multirow{2}{*}{\bf AVG} & 
        \multirow{2}{*}{\bf INC (\%)}\\\cline{1-3}
\bf Modifier &
	\bf Prop. &
	\bf Branch &
	&
	&
	&
        &
        \\\hline
\multirow{2}{*}{static non-final} &
	 \multirow{2}{*}{int} &
	 first switch case &
	2085 &
	86 &
	46 &
	69.5 &
	33.3\\\cline{3-8}
 &
	  &
	 switch default &
	1347 &
	61 &
	40 &
	44.9 &
	-4.7\\\hline
\multirow{2}{*}{static final} &
	 \multirow{2}{*}{int} &
	 first if path &
	1198 &
	56 &
	36 &
	39.9 &
	-23.4\\\cline{3-8}
 &
	  &
	 last else path &
	8973 &
	350 &
	282 &
	299.1 &
	534.6\\\hline
\multirow{2}{*}{static final} &
	 \multirow{2}{*}{String} &
	 first switch case &
	8550 &
	306 &
	280 &
	285.0 &
	446.7\\\cline{3-8}
 &
	  &
	 switch default &
	9637 &
	684 &
	303 &
	321.2 &
	581.6\\\hline
\multicolumn{8}{|c|}{\bf Baseline}\\\hline
\bf Modifier &
	\bf Prop. &
	\bf Branch &
        \bf SUM &
	  \bf MAX &
	\bf MIN &
	\bf AVG & 
        \bf INC (\%)\\\hline 
\multirow{2}{*}{static final} &
	 \multirow{2}{*}{int} &
	 first switch case &
	1564 &
	133 &
	47 &
	52.1 &
	0.0\\\cline{3-8}
 &
	  &
	 switch default &
	1414 &
	61 &
	45 &
	47.1 &
	0.0\\\hline
\end{tabular}
}
\end{table}

Although we used Java Development Kit 11 (\jdk 11) in all our experiments, we also ran the micro-experiment on \jdk 17 (Table~\ref{tab:benchmark-jdk17}) and 21 (Table~\ref{tab:benchmark-jdk21}) to observe if there were differences in more recent Java JDKs and if the results would hold.

For instance, considering \jdk 11 (Table~\ref{tab:benchmark-jdk11}), the last two lines correspond to the best combination (\texttt{static final}, \texttt{int}, \texttt{switch}) for the first and default case. In this variation, it takes 36.3ms to execute the code when the first case is selected and 38.1ms for the default case. Observe that by using the same kind of variable set by \texttt{setprop(debug.MUID)} but changing the \texttt{switch} by an \texttt{if} statement, we got a time improvement from 36.3ms to 34.6ms for the first branch (approximately 4.8\% better -- INC column) but, on the other hand, the \texttt{else} branch (or default branch) took 291.2ms to decide the mutant, against 38.1ms from the \texttt{switch} implementation default case, an increment of 664.3\%. We observed that the variation with the most significant performance penalty used a \texttt{String} instead of an \texttt{int} for the mutant identifier.

With some slight variance, we obtained the same results for \jdk 17 and \jdk 21. The most significant difference was that using a final variable did not bring as much improvement as in \jdk 11.

Taking these results into consideration, we decided to implement our schemata strategy by using \texttt{switch} statement controlled by a \texttt{static final int} variable, set by \texttt{adb setprop(debug.MUID)} command.

\Cref{fig:omni_original} shows an excerpt of code from one of the tested projects, and \Cref{fig:omni_schemata} shows the same code after applying the schemata strategy in a statement that is the target of three different mutations by three separate mutation operations.

\begin{figure}[!htb]
    \centering
\begin{minted}[frame=lines,fontsize=\scriptsize,samepage=true]{java}
android.content.Intent jobIntent;
jobIntent = new android.content
   .Intent(android.content.Intent.ACTION_SEND);
context.startService(jobIntent);
\end{minted}
 \caption{Original code (from project OmniNotes).}
    \label{fig:omni_original}
\end{figure}

The numbers in the \texttt{switch} cases do not increment linearly because they represent mutation IDs. We use a scheme to identify each mutation deterministically (even if parallelization is used). In this way, the same mutation has the same ID number regardless of whether it is generated by the mutant schemata or the traditional strategy.
This way, directly comparing the individual mutations between strategies is more straightforward.

\Cref{fig:muid} shows the current implementation of the method \texttt{getMUID()} that fetches the value of the property that will control which mutation is executed (if any). It creates a new process when the class is first loaded, which executes the command \textit{getprop debug.MUID}.
This command obtains the specific value associated with the \textit{debug.MUID} property.
%The created function \textit{getMUID()} can be seen next.

\begin{figure}[!htb]
    \centering
\begin{minted}[frame=lines,fontsize=\scriptsize,samepage=true]{java}
android.content.Intent jobIntent;
switch(MUID_STATIC) {
  // 0_NullIntentOperatorMutator
  case 27: {
    jobIntent = null;
    break;
  }
  // 1_InvalidKeyIntentOperatorMutator
  case 1027: {
    jobIntent = new android.content.Intent((Context) null,
      it.feio.android.omninotes.
        async.AlarmRestoreOnRebootService.class);
    break;
  }
  // 2_RandomActionIntentDefinitionOperatorMutator
  case 2027: {
     jobIntent = new android.content
         .Intent(android.content.Intent.ACTION_SEND);
       break;
  }
  default: {
    jobIntent = new android.content.Intent(context, 
      it.feio.android.omninotes.
        async.AlarmRestoreOnRebootService.class);
  break;
}
\end{minted}
 \caption{Code after applying the schemata strategy (from project OmniNotes).}
    \label{fig:omni_schemata}
\end{figure}

\begin{figure}[!htb]
    \centering
\begin{minted}[frame=lines,fontsize=\scriptsize,samepage=true]{java}
public static int getMUID() {
  String propertyValue = "-1";
  try {
    java.lang.Process process = 
      Runtime.getRuntime().exec("getprop debug.MUID");
    InputStream inputStream = process.getInputStream();
    BufferedReader reader = new BufferedReader(
      new InputStreamReader(inputStream));
    propertyValue = reader.readLine();
    reader.close();
    inputStream.close();
  } catch (IOException e) {
    Log.e("ERROR", String.valueOf(e));
  }
  return Integer.parseInt(propertyValue);
}
\end{minted}
 \caption{Method that fetches the property that controls the executed mutation.}
    \label{fig:muid}
\end{figure}

To simplify code generation, we add a copy of the function \texttt{getMUID()} and of the variable \texttt{MUID\_STATIC} to every class of the project. However, this means we execute the command \texttt{getprop} the first time every class is loaded to set the \texttt{MUID\_STATIC} variable.
This can incur significant overhead, which can be solved by having a single, possibly new, class that loads and stores the parameter and is statically accessed by all the other classes. This was left for future work.

\subsection{Architecture}

\begin{figure}[!htb]
\centering
\includegraphics[clip, trim=0.75cm 0.3cm 0.75cm 0.3cm,width=\columnwidth]{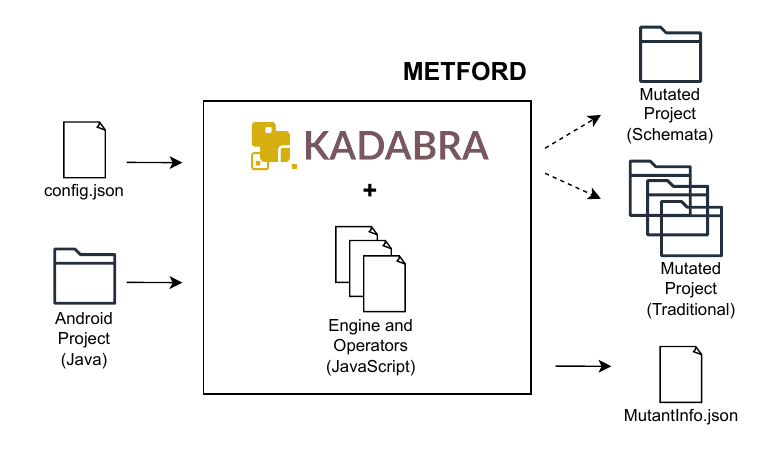}
\caption{METFORD flow}\label{fig:metford}
\end{figure}

Figure~\ref{fig:metford} illustrates the main flow of METFORD execution.
As input, it receives an Android project with Java code. The execution is controlled via JSON file (\texttt{config.json}), which contains several configuration options such as \texttt{classpath}, to specify compiled Java classes, or \texttt{operatorNameList}, to identify the mutation operators\footnote{A comprehensive list of supported options can be found in \url{https://github.com/specs-feup/mutation-testing-v2}} we intend to use.

Internally, METFORD uses the Kadabra\footnote{\url{https://github.com/specs-feup/kadabra}} Java source-to-source compiler to load the project Java files and execute the JavaScript code that encodes/implements the mutation operators.

The tool can output a single project with all mutations (mutant schemata strategy) or generate one project for each mutation (traditional mutation strategy). The input configuration file controls the type of output.
It also outputs a report in JSON format containing information about all applied mutations.

\revblue{Kadabra supports both mutation strategies. In Section~\ref{sec:rq1}, we show that both are equivalent in mutant behavior. Each mutant, independently of the adopted strategy, has a single ID, which makes it possible to use mutant schemata or traditional mutation simultaneously. 

Once the tester desires to visualize a single mutant, the tool uses the traditional mutation to show the mutant for the tester, avoiding the complexity of analyzing mutants in a file with a large \texttt{switch} statement. Mutant schemata can only speed up the mutant execution on mobile devices. Kadabra controls the mutation execution independently of the mutation strategy, avoiding mistakes during this step.

Moreover, it is important to observe that the chosen mutation strategy did not impact code coverage measurement once code coverage must be measured by executing the test set in the original source code and not in the mutated one.}

\subsubsection{Kadabra}

Kadabra is a source-to-source compiler based on the LARA framework, a Java library for building source-to-source compilers (also known as transpilers)~\cite{cardoso2016performance,pinto2018aspect}.

LARA-based compilers follow an architecture where analysis and transformation passes are implemented outside the compiler as JavaScript files (or, more recently, TypeScript).
They also provide access to an \ac{AST}-like structure that represents the input source code and which the JavaScript input scripts can inspect and manipulate. The \acp{AST} used in this framework supports generating the source code again (also known as ``unparse''), reflecting changes in the structure of the generated source code.

Each LARA-based compiler supports a single language~\cite{carvalho2023dsl} or family of languages~\cite{bispo2020clava}.
Kadabra~\cite{carvalho2023dsl} is a LARA-based source-to-source compiler for the Java language, which applies analyses and transformations over Java source code.
Under the hood, Kadabra uses as an internal representation of the code the \ac{AST} provided by Spoon, the open-source Java-to-Java library developed by the Spirals research team~\cite{https://doi.org/10.1002/spe.2346}.
We have implemented the mutant schemata strategy presented in this work as a collection of JavaScript files, which form a mutation testing library for Kadabra (see \Cref{fig:metford}).

\subsection{Mutation Operators}
\citet{SilvaSTVR22} present a mapping study on mutation testing for mobile applications. In his study, the mutation operators are divided into eight classes: configuration, connectivity, GUI, intent, location, persistence, sensors, and traditional. 

\revblue{We did not implement all 138 mutation operators identified by \citet{SilvaSTVR22}. Still, we have mutation operators of four of the eight classes grouped differently. We have implemented GUI, intent, sensors, and traditional mutation operators once they represent a large fault model for the apps under testing. The Android-specific mutant operators include operators related to GUI, intent, and sensors. \citet{SilvaSTVR22}'s traditional category was divided into General and Java-specific categories.

In the future, we intend to extend the set of mutation operators. Kadabra is an easy-to-extend tool. Mutation operators may be implemented by using the JavaScript mutation operator file descriptor. It allows new mutation operators to be added without recompiling the tool itself, just by creating new script files that are added to a specific folder and registered in a specific file. In this study, we intend not to implement all possible mutation operators but to explore the feasibility of the meta-mutant approach and compare it to the traditional approach using a set of relevant operators.} Our work focuses on intent, GUI, and traditional categories.

\begin{itemize}
    \item 
The Intent category encompasses changes related to the specific Android component Intent,
which is widely used in Android applications to enable communication and interaction
among different Android components such as Activities, Services, and Broadcast
receivers (see Android Specific mutants in highlight in bold in Table~\ref{tab:operators}).
    \item 
The GUI category includes changes that modify GUI elements, as the name suggests,
such as XML files, Activities, and event handlers (see Android Specific mutants highlight in bold in  Table~\ref{tab:operators}).
    \item 
The Traditional category introduces changes based on common errors in the Java programming
language (see General Specific highlight in bold in Table~\ref{tab:operators}).
\end{itemize}

\begin{table}[!htb]
\caption{Mutation Operators implemented.}
\label{tab:operators}
\centering
\resizebox{\columnwidth}{!}{%
\begin{tabular}{|c|c|c|}
\hline
\textbf{General Specific (GS)} &
  \textbf{Java Specific (JS)} &
  \textbf{Android Specific (AS)} \\ \hline
\begin{tabular}[c]{@{}c@{}}\textbf{Arithmetic}\\ Bitwise\\ Comparison\\ Logical\\ Assignment\\ Arithmetic Deletion\\ Bitwise Deletion\\ Comparison Deletion\\ Logical Deletion\\ Assignment Deletion\\ Constant\\ Unary\\ Unary Logical Negation\\ Unary Deletion\end{tabular} &
  \begin{tabular}[c]{@{}c@{}}Constructor Call\\ Remove Conditional\\ Non Void Call\\ Nullify Input Variable\\ Nullify Return Value\\ Return Value Operator\\ Invalid Date\\ Invalid Method Call Argument\\ Null Method Call Argument\\ Not Serializable\\ Fail On Null\\ String Argument Replacement\\ String Call Replacement\\ Conditional Expression Replacement\end{tabular} &
  \begin{tabular}[c]{@{}c@{}}\textbf{Buggy GUI Listener}\\ \textbf{Lengthy GUI Listener}\\ \textbf{Lengthy GUI Creation}\\ \textbf{Find View By Id Returns Null}\\ \textbf{View Component Not Visible}\\ \textbf{Invalid View Focus}\\ \textbf{Invalid ID FindView}\\ \textbf{Null Intent}\\ \textbf{Random Action Intent Definition}\\ \textbf{Intent Target Replacement}\\ \textbf{Invalid Key Intent}\\ \textbf{Null Value Intent PutExtra}\\ \textbf{Intent Payload Replacement}\\ XML Edit TextWidget Invisible\\ XML ViewGroup Widget Invisible\\ XML Button Widget Invisible\\ XML EditText Widget Deletion\\ XML Button Widget Deletion\\ XML TextView Widget Deletion\\ XML Invalid Color\\ XML Button Widget Change Appearance\\ XML EditTex tWidget Change Appearance\\ XML ViewGroup Widget Change Type\\ Sensor Null Bluetooth Adapter\\ Sensor Null GPS Location\end{tabular} \\ \hline
\end{tabular}%
}
\end{table}

Considering our intention of supporting mutation testing on Android Mobile Applications, from the list of mutation operators presented in Table~\ref{tab:operators}, we tested in Kadabra, one general specific operator (i.e., Arithmetic), and all Android Specific operators that are not XML (e.g., XML Invalid Color) or resource (e.g., Null GPS Location) related.

Table~\ref{tab:mutation-operators} contains all the mutation operators we used in our experiment. As observed, three mutation operators from Table~\ref{tab:mutation-operators} use a random number generator for the mutant generation. 

The Android Reference Documentation for Intent\footnote{\url{https://developer.android.com/reference/android/content/Intent}} class has several constants defining different intent actions. For instance, considering mutation operator 13 -- RandomActionIntentDefinition, the random number is used to pick one from a predefined list inside the operator definition to create the mutant. Figure~\ref{fig:random} illustrates a mutant of the RandomActionIntentDefinition mutation operator, which replaces the original \texttt{intent} (lines 15 to 18) with the mutated \texttt{intent} (lines 9 to 11).

\begin{table}[!htb]
\caption{Set of Mutation Operators Used in the Experiment.}\label{tab:mutation-operators}
\resizebox{\columnwidth}{!}{%
\begin{tabular}{|c|l|l|c|c|}\hline
\bf ID &
	\bf Mutation Operator &
        \bf Acronym &
	\bf Class &
	\bf Rand\\\hline
1 &
	BinaryMutator (Arithmetic) &
        BMA &
	GS &
	No\\\hline
2 &
	BuggyGUIListener &
        BGL &
	AS &
	No\\\hline
3 &
	FindViewByIdReturnsNull &
        FVBIRN &
	AS &
	No\\\hline
4 &
	IntentPayloadReplacement &
        IPLR & 
	AS &
	No\\\hline
5 &
	IntentTargetReplacement &
        ITR &
	AS &
	%21
        Yes\\\hline
6 &
	InvalidIDFindView &
        IIFV & 
	AS &
	%22
        Yes\\\hline
7 &
	InvalidKeyIntent &
        IKI &
	AS &
	No\\\hline
8 &
	InvalidViewFocus &
        IVF & 
	AS &
	No\\\hline
9 &
	LengthyGUICreation &
        LGC & 
	AS &
	No\\\hline
10 &
	LengthyGUIListener &
        LGL & 
	AS &
	No\\\hline
11 &
	NullIntent &
        NI &
	AS &
	No\\\hline
12 &
	NullValueIntentPutExtra &
        NVIPE & 
	AS &
	No\\\hline
13 &
	RandomActionIntentDefinition &
        RAID & 
	AS &
	%23
        Yes\\\hline
14 &
	ViewComponentNotVisible &
        VCNV & 
	AS &
	No\\\hline
\end{tabular}
}
\end{table}

\begin{figure}[!htb]
    \centering
\begin{minipage}{0.9\columnwidth}    
\begin{minted}[frame=lines,fontsize=\scriptsize,samepage=true,linenos]{java}
private void startEditEntryActivity(
 int requestCode, 
 com.beemdevelopment.aegis.vault.VaultEntry entry) {
 android.content.Intent intent;
 switch(MUID_STATIC) {
 ...
  // 38_RandomActionIntentDefinitionOperatorMutator
  case 38173: {
   intent = new android.content.Intent(
    android.content.Intent.ACTION_SEND
    );
    break;
  }
  default: {
    intent = new android.content.Intent(
     this, 
     com.beemdevelopment.aegis.ui.EditEntryActivity.class
    );
    break;
  }
 }
 ...
}
\end{minted}
\end{minipage}
 \caption{Code snippet of Aegis of method \texttt{startEditEntryActivity} inside \texttt{MainActivity} class,  illustrating a mutant of RandomActionIntentDefinitionOperatorMutator.}
    \label{fig:random}
\end{figure}

The last column indicates mutation operators with randomness characteristics. For these mutation operators, Kadabra configuration files provide a seed parameter such that the choice of random numbers becomes deterministic. This allows the use of the mutation operator on both strategies (Schemata and Traditional) and ensures the same random numbers will be used on both strategies.

\revblue{It is essential to notice that mutation testing in industrial settings demands prioritization, and we need to reduce the number of operators to make mutation testing feasible~\cite{Petrovic22PMTS}. Moreover, even with the current set of mutation operators, we observed the weakness of the currently available test sets in detecting faults modeled by these mutation operators.}

\subsection{Mutation Operators Implementation}

All mutation operators were implemented as individual and independent JavaScript scripts executed by the Kadabra compiler.
All operators implement the interface \texttt{Mutator} and its methods, the most relevant for this work being:

\begin{itemize}
    \item \texttt{.addJp(\$jp)}: adds a point in the code to the mutation operator, internally this can result in zero or more mutations ready to be applied in that point.
    \item \texttt{.hasMutations()}:  true if the mutation operator still has mutations to apply, false if all possible mutations were applied.
    \item \texttt{.mutate()}: applies the next mutation. 
    \item \texttt{.getMutationPoint()}: returns the point in the code with the last mutation.     
    \item \texttt{.restore()}: reverts the changes introduced by the mutation.
\end{itemize}

The interface follows a pattern similar to an iterator, and its most noticeable feature is the presence of a \texttt{restore()} method.
This interface predates the mutant schemata strategy and was initially used to implement the traditional mutation strategy.
In this context, the \texttt{restore()} method was a way to improve the performance of the traditional mutant generation. Instead of parsing the code anew for each mutation, one only needed to revert the previous mutation directly in the \ac{AST} (which usually is a very fast operation, compared with parsing the code) after generating the mutated code, leaving the \ac{AST} ready for applying the next mutation.

Coincidentally, this same interface can be used to implement the schemata strategy, allowing the reuse of mutation operators between strategies. 

To create a new mutation operator, two steps are required: adding a new script to the mutation operators folder with a class that implements the \texttt{Mutator} interface and registering the new operator in a script that contains the list of available operators. 
This allows for a lightweight development and debugging cycle and the easy addition of new operators.

\subsection{Mutants Execution}

Since our approach is based on the Kadabra compiler, to generate the mutants, one only needs to install \jdk 17 or later to execute the tool.

Generating code for real medium to big-sized Java projects always has challenges; this was no exception. The underlying library that Kadabra uses, Spoon, had trouble generating certain types of code, particularly inner classes, which we solved by introducing two mechanisms: patching and a class exclude list. 

Since Kadabra interprets arbitrary JavaScript code that can change the \ac{AST}, we can specify a custom script (i.e., with the option \texttt{patch}) which can be executed before we apply the mutation.
This script can change the code as required and target specific files and methods.
Examples of changes we have applied include removing some \texttt{final} modifiers or setting the correct fully qualified name of certain \texttt{this} references.

As the name indicates, the class exclude list (i.e., with the option \texttt{excludeList}) allows to specify classes that should not be targeted. In this case, METFORD will use the original file in the new version of the project, avoiding generating a file with errors. 

After execution, METFORD generates a report in JSON format (i.e., \texttt{MutationInfo.json}), which contains structured information about each applied mutation, the file and line in which it was applied, and its arguments.
Unique identifiers are generated for each mutation, corresponding to the same mutation number in the traditional and schemata approaches. 

\subsubsection{Mutant Execution in Mobile Applications}

Mutation testing in traditional software products is challenging. Mobile applications introduce additional issues, mainly because of the environment required to run the application and its mutants. Figures~\ref{algo:schemata} and~\ref{algo:traditional} present the basic algorithms we used to run Mutant Schemata and Traditional mutants, respectively.

\revred{We run each mutated app version on each strategy (schemata or traditional) only once with all test cases. When we found a discrepancy in a specific test case, we reran only that test case in the particular app on both strategies to identify if this discrepancy was due to behavior differences in the mutation strategy or flakiness. All identified discrepancies were due to flakiness.}

As can be observed, the main difference between mutation strategies is that in Mutant Schemata we compile and build the original application and all mutated variations (the meta mutant) in a single step (line 2 of the algorithm in Figure~\ref{algo:schemata}). We push and install the built app only once, independently of the number of mutants, once all of them are embedded in the compiled source (line 3 of the algorithm in Figure~\ref{algo:schemata}). 

\begin{algorithm}[!htb]
\scriptsize
\DontPrintSemicolon
\hrule
\KwIn{A set of identifiers $ID = \{ id_1, id_2, \ldots, id_m \}$}
\KwIn{A set of test cases $T = \{ tc_1, tc_2, \ldots, tc_n \}$}
\KwOut{CSV file with the execution time step-by-step}
\KwOut{CSV file with the killing matrix}
Start emulator {\tt --wipe-data}\\
\colorbox{lightgray}{Compile and Build Schemata App apk}\\ 
\colorbox{lightgray}{Push and Install Schemata App apk}\\
Compile and Build Android Test apk\\
Push and Install Android Test apk\\

\ForEach{$id \in ID$}{%
    \uIf{ emulator {\bf not} active }{%
        Restart {\it emulator}
    }
    \colorbox{lightgray}{Set active {\tt debug.MUID} to $id$}\\
    \ForEach{$tc \in TC$}{%
        \uIf{ emulator {\bf not} active }{%
            Restart {\it emulator}\\
		\colorbox{lightgray}{Set active {\tt debug.MUID} to $id$}\\
        }
        Run $tc$ on Mutant $id$\\
        Record status of $tc$ for Mutant $id$
    }
}
Stop emulator\\
\textbf{Close} CSV file with the execution time\\
\textbf{Close} CSV file with the killing matrix
\hrule

\caption{Algorithm for schemata execution.}
\label{algo:schemata}
\end{algorithm}

\begin{algorithm}[!htb]
%\sffamily
\scriptsize
\DontPrintSemicolon
\hrule
\KwIn{A set of identifiers $ID = \{ id_1, id_2, \ldots, id_m \}$}
\KwIn{A set of test cases $T = \{ tc_1, tc_2, \ldots, tc_n \}$}
\KwOut{CSV file with execution time step-by-step}
\KwOut{CSV file with the killing matrix}
Start emulator {\tt --wipe-data}\\
Compile and Build Android Test apk\\
Push and Install Android Test apk\\
\ForEach{$id \in ID$}{%
    \uIf{ emulator {\bf not} active }{%
        Restart {\it emulator}
    }
    \colorbox{lightgray}{Replace Mutant $id$ source code}\\
    \colorbox{lightgray}{Compile and Build Mutant $id$ apk}\\
    \colorbox{lightgray}{Push and Install Mutant $id$  apk}\\
    \ForEach{$tc \in TC$}{%
        \uIf{ emulator {\bf not} active }{%
            Restart {\it emulator}\\
	}
        Run $tc$ on Mutant $id$\\
        Record status of $tc$ for Mutant $id$
    }
}
Stop emulator\\
\textbf{Close} CSV file with the execution time\\
\textbf{Close} CSV file with the killing matrix
\hrule

\caption{Algorithm for traditional execution.}
\label{algo:traditional}
\end{algorithm}

Later, to select a specific mutant for execution, we set a property \texttt{debug.MUID} in our Bash script by using the following command:\\[3pt]

\begin{minipage}{0.9\columnwidth}
\begin{minted}[frame=lines,fontsize=\scriptsize,samepage=true,linenos]{bash}
adb shell setprop debug.MUID "${id}"
\end{minted}
\end{minipage}\\[3pt]

On the other hand, when using the traditional mutation strategy, illustrated in the algorithm in Figure~\ref{algo:traditional}, we do not need to set a property to select a given mutant because we have a different file for each mutant. Nevertheless, we must replace the previous source code with the new one, perform the compile and build process before each new mutant's execution, and push and install the generated APK into the device before the test set run (lines 7 to 9 in the algorithm in Figure~\ref{algo:traditional}).

\subsubsection*{Execution time computation}

To measure the time, we use the Bash \texttt{date} command, as illustrated below, where \verb+<<LABEL>>+ is replaced by a BEGIN and END tag before and after each command of interest.\\[3pt]

\begin{minipage}{0.9\columnwidth}
\begin{minted}[frame=lines,fontsize=\scriptsize,samepage=true,linenos]{bash}
date --rfc-3339=ns |tr -d "\n" &>> time.csv
echo ";<<LABEL>>" &>> time.csv
\end{minted}
\end{minipage}\\[3pt]

Consider Figure~\ref{fig:time-report-schemata}, which contains a piece of the time report for mutant schemata execution, and Figure~\ref{fig:time-report-traditional}, which presents the same time report for the traditional execution. In our experiment, an $id$ with value -1 means the original program, and all other $id$ values with positive numbers identify a specific mutant. For schemata, \texttt{BEGIN} and \texttt{END} tags also include the time to set \texttt{debug.MUID} property before running tests. For traditional, between \texttt{BEGIN} and \texttt{END} tags includes the time for copying, compiling, building, pushing, and installing the mutant before running tests.

\begin{figure}[!htb]
    \centering
\begin{minipage}{0.9\columnwidth}
\begin{minted}[frame=lines,fontsize=\tiny,samepage=true,linenos]{text}    
2024-03-26 17:50:10.839406062;START TIME
2024-03-26 17:50:10.840827605;START EMULATOR
2024-03-26 17:50:55.855212902;BEGIN BUILD APP AND TEST APK
2024-03-26 17:51:38.875542330;END BUILD APP AND TEST APK
2024-03-26 17:51:38.877918660;BEGIN INSTALL APP AND TEST APK
2024-03-26 17:51:39.745081721;END INSTALL APP AND TEST APK
2024-03-26 17:51:39.755937908;BEGIN RUN MUTANT -1
2024-03-26 17:54:33.777333458;END RUN MUTANT -1
2024-03-26 17:54:33.780446872;BEGIN RUN MUTANT 6
2024-03-26 17:57:13.840030662;END RUN MUTANT 6
...
2024-03-28 06:20:50.348967575;STOP EMULATOR
2024-03-28 06:21:00.351141029;END TIME
\end{minted}
\end{minipage}
    \caption{Sample of time report for Schemata Execution}
    \label{fig:time-report-schemata}
\end{figure}

\begin{figure}[!htb]
\centering
\begin{minipage}{0.9\columnwidth}
\begin{minted}[frame=lines,fontsize=\tiny,samepage=true,linenos]{text}
2024-03-28 06:21:00.410311859;START TIME
2024-03-28 06:21:00.412316770;STARTING EMULATOR
2024-03-28 06:21:45.423332009;BEGIN BUILD APP AND TEST APK
2024-03-28 06:22:52.598934588;END BUILD APP AND TEST APK
2024-03-28 06:22:52.600044333;BEGIN INSTALL TEST APK
2024-03-28 06:22:52.708210939;END INSTALL TEST APK
2024-03-28 06:22:52.720756067;BEGIN RUN MUTANT -1
2024-03-28 06:22:52.735212873;BEGIN BUILD MUTANT APK -1
2024-03-28 06:22:54.461427329;END BUILD MUTANT APK -1
2024-03-28 06:22:54.462342253;BEGIN INSTALL MUTANT APK -1
2024-03-28 06:22:54.598058951;END INSTALL MUTANT APK -1
2024-03-28 06:25:42.746910106;END RUN MUTANT -1
2024-03-28 06:25:42.754841191;BEGIN RUN MUTANT 6
2024-03-28 06:25:42.764396035;BEGIN BUILD MUTANT APK 6
2024-03-28 06:25:47.114631889;END BUILD MUTANT APK 6
2024-03-28 06:25:47.115556190;BEGIN INSTALL MUTANT APK 6
2024-03-28 06:25:47.247326797;END INSTALL MUTANT APK 6
2024-03-28 06:28:32.849185780;END RUN MUTANT 6
...
2024-03-29 21:14:56.212598599;STOP EMULATOR
2024-03-29 21:15:06.218927958;END TIME
\end{minted}
\end{minipage}
\caption{Sample of time report for Traditional Execution}
\label{fig:time-report-traditional}
\end{figure}

\subsubsection*{Test Execution Alternatives}

Our scripts, available at our GitHub repository\footnote{\url{https://github.com/specs-feup/paper-metford}}, allow two ways of running test cases against mutants: 
\begin{enumerate}
    \item All against all~\cite{https://doi.org/10.1002/stvr.1769}: must run all test cases against all mutants, even if a given mutant was killed by a previous test case. This allows the creation of the kill matrix, which is helpful for experimental purposes, especially for considering the definition of test set minimization or minimal mutants identification~\cite{Delamaro18WMMU}. In our scripts, this is identified by the \texttt{fullFail} parameter.

    \item All against mutants remaining alive~\cite{https://doi.org/10.1002/stvr.1769}: when a test case kills a mutant, we stop running tests on that mutant, i.e., we do not run the remaining tests of the test suite.
    With this approach, depending on the order in which we execute the test cases, we may speed up the execution time, but we cannot build the entire killing matrix. Also, we may need the all-against-all approach to find the best test set order and use it in the following test case executions. In our scripts, this is identified by the \texttt{fastFail} parameter.
\end{enumerate}

\subsubsection*{Test Timeout}

Finally, when we create a mutant, we cannot infer its behavior when executed. For example, we may create a mutant that causes infinite looping during execution. Although \emph{x}Unit testing frameworks like JUnit\footnote{\url{https://junit.org/junit5/docs/current/api/org.junit.jupiter.api/org/junit/jupiter/api/Timeout.html}} may offer options to set test case execution timeouts, we cannot guarantee that all test cases available with software programs use such a feature. 

In this sense, we implemented a timeout mechanism in our scripts using the Bash \texttt{timeout} command to avoid checking and changing the test set code to support such a feature. Figure~\ref{fig:script-timeout} illustrates a piece of the script code responsible for controlling the test case execution timeout.

Observe the \texttt{cmd} definition on lines 14 and 16. It makes use of the \texttt{timeout} command to set a time limit (\texttt{TIMEOUT}) to run the \verb+adb shell+ command, responsible for running a specific test case \texttt{tc} by using a test case executor identified by \texttt{TCEXECUTOR}.

Following, in lines 20 and 21, we run \texttt{cmd} command preceded by \texttt{time}, which registers the time spent on running each test case \texttt{tc} (\texttt{tc.time}), and its output (\texttt{tc.out}). 

Then, we check the return status of the \texttt{cmd} command. If  \texttt{cmd} ends by timeout, it returns  124, and sets the \texttt{status} variable to 124 (line 25). On the other hand, if it does not end due to timeout, we recover test case execution status from \texttt{tc.out} (line 27). Finally, we set the status in the killing matrix (line 29) and, depending on the status and execution alternative (\texttt{fullFail} or \texttt{fastFail}), we may interrupt this mutant execution (lines 31 to 34) and go to next mutant, or continue with the same mutant but running another test case.

\begin{figure}[!htb]
\centering
\begin{minipage}{0.9\columnwidth}
\begin{minted}[frame=lines,fontsize=\scriptsize,samepage=true,linenos]{bash}
...
OUT="<<output-directory>>"
TIMEOUT=2m
TCEXECUTOR="<<test case executor>>"
for m in $(cat mutant-ids.txt)
do
  ...
  echo -n "$m" >> $OUT/killing-matrix.csv

  for tc in $(cat test-case-names.txt)
  do
    ...
    
    cmd="timeout ${TIMEOUT} adb shell am instrument \
         -w -r --no-window-animation \
         -e class $tc $TCEXECUTOR"
    
    cmdtime=$(date)
    
    /usr/bin/time -o $OUT/$m/$tc.time --quiet -v \
    -p ${cmd} >& ${OUT}/${m}/$tc.out

    if [ "$?" -eq 124 ]
    then
      status=124
    else
      status="<<get status from adb execution output>>"
    fi
    echo -n ";$status" >> $OUT/killing-matrix.csv

    if [[ "$STOP" == "fastFail" && $status -ne 0 ]]
    then
      break
    fi
  done
  ...
  echo "" >> $OUT/killing-matrix.csv
done
...
\end{minted}
\end{minipage}
 \caption{Test case timeout execution control example.}
    \label{fig:script-timeout}
\end{figure}

\revblue{We also have scripts allowing testers to control when the device/emulator can be restarted. They may choose to run all mutants and all test cases without restarting the emulator/device, or they may decide to restart the emulator/device before each mutant execution or before each test case execution on each mutant. These proposed mutation execution mechanisms make it possible to avoid inadvertently sharing states or resources, leading to unpredictable behavior during test execution. To speed up mutant execution, we only restart the emulator when it is not available to run the test case on a given mutant. In future work, we intend to investigate the impact of each mechanism during mutant execution.}

\section{Case Study}\label{sec:case-study}

The goal of these case studies is to be able to answer the following research questions:
\begin{itemize}
    \item RQ1: Do mutant schemata and traditional mutation strategies implementations exhibit the same behavior?; 

    \item RQ2: How do time and memory consumption compare in the two implementations (mutant schemata and traditional)?
\end{itemize}

\subsection{Subjects}\label{sec:subjects}

The selection process of the subjects followed eight requirements:
\begin{itemize}
    \item Android native application;
    \item Source code must be available;
    \item Written mainly in Java;
    \item The application should have test cases with Android Instrumented Tests\footnote{Android Instrumented Tests are the tests designed to run on Android physical or emulators devices and leverage the features and components of Android API, without the need for mocks or third-party libraries (e.g., Roboletric).}, and the number of test cases must exceed 20;
    \item The application should have at least one activity file;
    \item The application should use Gradle as the build tool to facilitate running and testing the application;
    \item The application must have had a release in the last two years;
    \item The application should have over 250 forks and 1000k stars at GitHub.
\end{itemize}

\revred{The application selection process started by browsing the Android projects on GitHub. Firstly, we searched for Android apps written in Java and filtered considering the number of forks ($\ge$ 250) and stars ($\ge$ 1000 stars) and the last release's date. For each application, we manually evaluated whether the app was mainly written in Java by defining a threshold of 55\%. After defining a set of candidates, we checked if the application contained Android instrumented tests by manually analyzing the ``androidTest'' folder and discarding those where the folder was empty. Next, each candidate application was cloned into a local machine and compiled using \jdk 11. The apps that had some build issues, such as the \textit{build failed} error message, \textit{gradle sync failed}, \textit{gradle version not supported}, and \textit{missing settings} were promptly discarded. Moreover, the applications were standardized using a gradle version compatible with \jdk 11 and the \textit{targetSdkVersion} (version 33). Additionally, we remove some building options (i.e., release options, multiple product flavors, and keystore information) to facilitate carrying out the experimental study.}

We selected six applications that fulfilled the previous criteria and could compile successfully. Table~\ref{tab:selapps} presents general information about the selected Android apps, such as the application ID, name, the number of Android instrumented tests (Total, Ignored, and Used), the year of the last application update, and the number of forks and stars. It is worth noting that the selected applications were also used in different studies~\cite{Dong21FTDAVEOE, Xiong23AESFBAA, Pan20RLBCDTAA, Su21BAGTAARWB}.

\begin{table}[!htb]
\caption{General information about the selected Android apps. }
\label{tab:selapps}
\centering
\resizebox{\columnwidth}{!}{%
\begin{tabular}{|l|l|r|r|r|r|r|r|r|}
\hline
\multicolumn{2}{|c|}{\textbf{App}} & 
    \multirow{2}{*}{\textbf{Version}} & 
    \multicolumn{3}{|c|}{Android Tests} & 
    \textbf{Last} & 
    \textbf{\#} & 
    \textbf{\#} \\ \cline{1-2}\cline{4-6}
\textbf{ID} & 
    \textbf{Name} &
     &
    \textbf{Tot.} &
    \textbf{Ign.} &
    \textbf{Used} &
    \textbf{Updated} &
    \textbf{Forks} & 
    \textbf{Stars} \\ \hline
AE &
    Aegis & 2.1.3 & 23 & 0 & 23 & 2024 & 338 & 7975 \\ \hline
    %release date: Feb 27, 2023
AF &
    AmazeFileManager & 3.8.4 & 36 & 5 & 31 & 2024 & 1531 & 5054 \\ \hline
    %release date: Nov 20, 2022
AP &
    AntennaPod & 3.3.2 & 103 & 7 & 96 & 2024 & 1321 & 5741 \\ \hline
    %release date: Mar 24, 2024
KP &
    KeePassDroid & 2.6.8 & 119 & 0 & 119 & 2022 & 347 & 1368 \\ \hline
    %release date: Oct 7, 2022
ON &
    Omni-Notes & 6.2.9 & 102 & 3 & 99 & 2023 & 1106 & 2663 \\ \hline
    %release date: Jun 1, 2023
SN &
    Simplenote & 2.30 & 74 & 6 & 68 & 2024 & 297 & 1722 \\ \hline
    %release date: Oct 11, 2023
\end{tabular}
}
\end{table}

Concerning the Android instrumented tests, we do not intend to fix them. In this sense, to minimize test flakiness~\cite{Ngo23RTFU,Romano21EAUB,Memon13ATGA}, we run all Android instrumented tests for each app on a set of 8 different emulators (Pixel 6 Pro with Android API 30 and 31, and Pixel 7 Pro with Android API 29, 30, 31, 32, 33, and 34). We ran each test 10 times per emulator, resulting in 80 test set runs per app. If a test case fails in one of those executions, it is ignored using the \verb+@Ignore+ JUnit annotation. Therefore, columns 4 to 6 of Table~\ref{tab:selapps} present data concerning the total, ignored, and effectively used Android instrumented test cases per app.

Following is a brief description of the applications used in the study.

\begin{itemize}
    \item Aegis (AE) Authenticator is a free, secure, open-source two-factor authenticator Android application\footnote{https://github.com/beemdevelopment/Aegis}.

    \item AmazeFileManager (AF) is an open-source file management program for Android, offering extensive directory access and various file-related functions\footnote{https://github.
com/TeamAmaze/AmazeFileManager}.

    \item AntennaPod (AP) is a podcast management and playback application offering immediate access to an extensive library of both free and premium podcasts\footnote{https://github.com/
AntennaPod/AntennaPod}.

    \item KeePassDroid (KP) is an open-source Android application for password management\footnote{https://github.com/bpellin/keepassdroid}.

    \item Omni-Notes (ON) is a user-friendly, open-source note-taking application\footnote{https://github.
com/federicoiosue/Omni-Notes}. 

   \item Simplenote (SN) is an Android application for note-taking and task list management\footnote{https://github.com/Automattic/simplenote-android}. 
\end{itemize}

Next, Table~\ref{tab:infoMutatedFiles} presents additional information about the selected apps regarding the files considered in the mutant generation phase. 

Column App identifies the app, and column \#M presents the number of sub-modules of the application. Column \#MM refers to the sub-modules used in the mutant generation phase. Despite some applications having more than one module (i.e., AntennaPod, AmazeFileManager, Simplenote), the mutants were generated only for those modules with instrumented tests. For instance, Simplenote (SN) app has modules \textit{Simplenote} and \textit{Wear}; however, only \textit{Simplenote} module has instrumented tests, and hence, we did not generate mutants for \textit{Wear} module.

\begin{table}[!htb]
\caption{General information about the mutated apps.}
\label{tab:infoMutatedFiles}
\centering
\resizebox{\columnwidth}{!}{%
\begin{tabular}{|l|r|r|r|r|r|r|r|}
\hline
\textbf{App} & \multicolumn{1}{l|}{\textbf{\#M}} & \multicolumn{1}{l|}{\textbf{\#MM}} & \multicolumn{1}{l|}{\textbf{\#CF}} & \multicolumn{1}{l|}{\textbf{\#JF}} & \multicolumn{1}{l|}{\textbf{\#KF}} & \multicolumn{1}{l|}{\textbf{\#JFMM}} & \multicolumn{1}{l|}{\textbf{\#JFM}} \\ \hline
AE & 1 & 1 & 188 & 188 & 0 & 188 & 74 \\ \hline
AP & 9 & 1 & 447 & 447 & 0 & 164 & 114 \\ \hline
AF & 3 & 1 & 404 & 200 & 202 & 163 & 68 \\ \hline
KP & 1 & 1 & 228 & 228 & 0 & 215 & 62 \\ \hline
ON & 1 & 1 & 168 & 162 & 6 & 162 & 47 \\ \hline
SN & 2 & 1 & 142 & 107 & 0 & 106 & 52 \\ \hline
\end{tabular}
}
\end{table}

Next, column \#CF refers to the total class files of the application considering the entire application, and columns \#JF and \#KF present the total Java and Kotlin class files, respectively. Finally, \#JFMM presents the total of candidates' Java class files of the mutated module, and \#JFM refers to the total of mutated Java class files. For instance, module \textit{Simplenote (SN)} has 106 Java class files (column \#JFMM) that could be mutated, and Kadabra generated mutants for 52 of those Java files (column \#JFM).

In~\cite{https://doi.org/10.1002/stvr.1769}, Mutant Schemata and Traditional strategies, mutant generation times were considered equivalent. In our case, Schemata generates the same set of mutants, on average, 8.5\% faster than traditional mutants, as presented in Table~\ref{tab:generation-time}. 

\begin{table}[!htb]
\caption{Schemata \emph{versus} Traditional Mutants Generation Time (seconds)}
\label{tab:generation-time}
\centering\scriptsize
%\resizebox{\columnwidth}{!}{%
\begin{tabular}{|l|r|r|r|r|}\hline
	\multirow{2}{*}{\bf App} &
	\bf Schemata &
	\bf Traditional &
	\bf \% Saving & 
        \bf \# Gen.\\
	 &
	 \bf Gen. Time &
	\bf Gen. Time &
	\bf Time &
        \bf Mutants\\\hline
AE &
	287.38 &
	316.83 &
	9.30 &
	  984\\\hline
AF &
	359.99 &
	386.77 &
	6.92 &
        1040\\\hline
AP &
	347.13 &
	368.88 &
	5.90 &
        1216\\\hline
KP &
	281.98 &
	286.56 &
	1.60 &
        406\\\hline
ON &
	160.92 &
	191.42 &
	15.93 &
        321\\\hline
SN &
	206.92 &
	233.37 &
	11.33 & 
        804\\\hline
\multicolumn{1}{|l|}{AVG} &
	274.05 &
	297.31 &
	8.50 & 
        795.17\\\hline
\multicolumn{1}{|l|}{SD} &
	77.81 &
	76.06 &
	4.92 & 
        360.30\\\hline
\end{tabular}%
%}
\end{table}

The last column of Table~\ref{tab:generation-time} contains the total number of generated mutants. As observed, the time is not constant regarding the number of mutants. For instance, AE and KP took almost the same time to generate Schemata mutants, but AE has more than twice the number of KP mutants. This inequality may be due to the set of operators that were applied to the apps and the differences in the implementation of the mutation operators, and this must be investigated in further experiments.

Also, considering the disk space required to store the generated mutants, comparing the original program with Schemata, from Table~\ref{tab:space}, we observed that Mutant Schemata correspond to an increment in original size around 44.77\% on average. On the other hand, the increment in disk space required to store each individual mutated project (column \% Inc below Traditional) is superior to 79,628\% in size. In this way, we observed that Mutant Schemata saves more than 99\% of disk space compared to Traditional. 

\begin{table}[!htb]
\caption{Schemata \emph{versus} Traditional Space (kbytes)}
\label{tab:space}
\centering
\resizebox{\columnwidth}{!}{%
\begin{tabular}{|l|r|r|r|r|r|r|}\hline
\multirow{2}{*}{\bf App} &
	\bf Original &
	\multicolumn{2}{|c|}{\bf Schemata} &
	\multicolumn{2}{|c|}{\bf Traditional} &
	\bf \% Saving \\\cline{3-6}
 &
	\bf Size &
	\bf Size &
	\bf \% Inc &
	\bf Size &
	\bf \% Inc &
	\bf Size\\\hline
AE &
	1364 &
	2048 &
	50.15 &
	1342228 &
	98303.81 &
	99.85\\\hline
AF &
	3404 &
	4332 &
	27.26 &
	3549368 &
	104170.51 &
	99.88\\\hline
AP &
	1444 &
	2324 &
	60.94 &
	1761696 &
	121901.11 &
	99.87\\\hline
KP &
	1524 &
	1984 &
	30.18 &
	620228 &
	40597.38 &
	99.68\\\hline
ON &
	1232 &
	1604 &
	30.19 &
	396536 &
	32086.36 &
	99.60\\\hline
SN &
	1248 &
	2120 &
	69.87 &
	1008504 &
	80709.62 &
	99.79\\\hline
\multicolumn{1}{|l|}{AVG} &
	1702.67 &
	2402.00 &
	44.77 &
	1446426.67 &
	79628.13 &
	99.78\\\hline
\multicolumn{1}{|l|}{SD} &
	840.99 &
	974.41 &
	18.18 &
	1140987.75 &
	36118.98 &
	0.11\\\hline
\end{tabular}%
}
\end{table}

\subsection{Answer to RQ1}\label{sec:rq1}

Previously, we tested whether the implementation of each mutation operator was correct on code snippets.  However, to be more confident in the implementation correctness, we aimed to test the mutation operators' implementation over real apps. So, we used the six apps presented in Table~\ref{tab:selapps} and all the mutation operators presented in Table~\ref{tab:mutation-operators}. We generated mutants based on two strategies:  Traditional (TR: one file per mutant) and Mutant Schema (SC: one file for all mutants). The results are in Table~ \ref{tab:comparison}. Each mutation operator generated the same number of mutants independently of the mutation type, Traditional or Schemata.

\begin{table}[!htb]
\caption{Total number and percentage of generated mutants by each mutation operator per app}
\label{tab:comparison}
\centering
\resizebox{\columnwidth}{!}{%
\begin{tabular}{|l|r|r|r|r|r|r|r|r|r|r|r|r|r|r|}\hline
\multirow{2}{*}{Operator} &
	\multicolumn{2}{|c|}{AE} &
	\multicolumn{2}{|c|}{AF} &
	\multicolumn{2}{|c|}{AP} &
	\multicolumn{2}{|c|}{KP} &
	\multicolumn{2}{|c|}{ON} &
	\multicolumn{2}{|c|}{SN} &
	\multicolumn{2}{|c|}{Total}\\\cline{2-15}
 &
	\# &
	\% &
	\# &
	\% &
	\# &
	\% &
	\# &
	\% &
	\# &
	\% &
	\# &
	\% &
	\# &
	\%\\\hline
BMA &
	225 &
	22,9 &
	326 &
	31,3 &
	321 &
	26,4 &
	132 &
	32,5 &
	71 &
	22,1 &
	114 &
	14,2 &
	1189 &
	24,9\\\hline
BGL &
	97 &
	9,9 &
	45 &
	4,3 &
	186 &
	15,3 &
	14 &
	3,4 &
	18 &
	5,6 &
	50 &
	6,2 &
	410 &
	8,6\\\hline
FVBIRN &
	59 &
	6,0 &
	28 &
	2,7 &
	15 &
	1,2 &
	5 &
	1,2 &
	11 &
	3,4 &
	13 &
	1,6 &
	131 &
	2,7\\\hline
IPR &
	36 &
	3,7 &
	37 &
	3,6 &
	14 &
	1,2 &
	6 &
	1,5 &
	14 &
	4,4 &
	53 &
	6,6 &
	160 &
	3,4\\\hline
ITR &
	0 &
	0,0 &
	0 &
	0,0 &
	0 &
	0,0 &
	0 &
	0,0 &
	0 &
	0,0 &
	0 &
	0,0 &
	0 &
	0,0\\\hline
IIFV &
	68 &
	6,9 &
	34 &
	3,3 &
	33 &
	2,7 &
	5 &
	1,2 &
	16 &
	5,0 &
	15 &
	1,9 &
	171 &
	3,6\\\hline
IKI &
	30 &
	3,0 &
	41 &
	3,9 &
	19 &
	1,6 &
	13 &
	3,2 &
	15 &
	4,7 &
	44 &
	5,5 &
	162 &
	3,4\\\hline
IVF &
	139 &
	14,1 &
	145 &
	13,9 &
	258 &
	21,2 &
	75 &
	18,5 &
	33 &
	10,3 &
	104 &
	12,9 &
	754 &
	15,8\\\hline
LGK &
	17 &
	1,7 &
	18 &
	1,7 &
	20 &
	1,6 &
	16 &
	3,9 &
	12 &
	3,7 &
	12 &
	1,5 &
	95 &
	2,0\\\hline
LGL &
	2 &
	0,2 &
	0 &
	0,0 &
	1 &
	0,1 &
	17 &
	4,2 &
	4 &
	1,2 &
	45 &
	5,6 &
	69 &
	1,4\\\hline
NI &
	28 &
	2,8 &
	40 &
	3,8 &
	15 &
	1,2 &
	10 &
	2,5 &
	18 &
	5,6 &
	42 &
	5,2 &
	153 &
	3,2\\\hline
NVIPE &
	36 &
	3,7 &
	37 &
	3,6 &
	14 &
	1,2 &
	6 &
	1,5 &
	14 &
	4,4 &
	53 &
	6,6 &
	160 &
	3,4\\\hline
RAID &
	108 &
	11,0 &
	144 &
	13,8 &
	62 &
	5,1 &
	32 &
	7,9 &
	62 &
	19,3 &
	155 &
	19,3 &
	563 &
	11,8\\\hline
VCNV&
	139 &
	14,1 &
	145 &
	13,9 &
	258 &
	21,2 &
	75 &
	18,5 &
	33 &
	10,3 &
	104 &
	12,9 &
	754 &
	15,8\\\hline
Total &
	984 &
	100,0 &
	1040 &
	100,0 &
	1216 &
	100,0 &
	406 &
	100,0 &
	321 &
	100,0 &
	804 &
	100,0 &
	4771 &
	100,0\\\hline

\end{tabular}
}
\end{table}

Observe from Table~\ref{tab:comparison} that, independently of the app, ITR (IntentTargetReplacementOperatorMutator) mutation operator did not generate mutants. Also, except for SimpleNote (SN), the BMA (BinaryMutatorArithmetic) is the one that produces more mutants; on average, 24.9\% of all mutants are of BMA. For SN, the operator RAID (RandomActionIntentDefinitionOperatorMutator) is the one that generates more mutants (19.3\% of SN mutants), followed by BMA (14,2\% of SN mutants). These differences in the percentage of generated mutants per app are interesting and show that each one has some particularities and different syntactic structures, which make it possible to have different percentages of mutants of each type in different apps. 

Following, we performed a dynamic evaluation by running the same Android instrumented test set against the complete set of mutants generated by both strategies: traditional and mutant schemata. Table~\ref{tab:score} presents the mutation score obtained on each mutation strategy for each app.

\begin{table}[!htb]
\caption{Flakiness impact on mutation score: Schemata versus Traditional.}
\label{tab:score}
\centering
\resizebox{\columnwidth}{!}{%
\begin{tabular}{|l|l|r|r|r|r|r|r|r|}\hline
\multirow{2}{*}{\bf App} &
	\multirow{2}{*}{\bf Strategy} &
	\multicolumn{3}{|c|}{\bf Mutants} &
	\bf Mutation &
	\multirow{2}{*}{\bf Difference} &
	\multirow{2}{*}{\bf Divergence} &
	\bf Divergence/\\\cline{3-5}
 &
	 &
	\bf Generated &
	\bf Alive &
	\bf Dead &
	\bf Score &
	 &
	 &
	\bf Generated\\\hline

\multirow{2}{*}{AE} &
	SC &
	984 &
	677 &
	  307$^*$ &
	31.2$^*$ &
	\multirow{2}{*}{11} &
	\multirow{2}{*}{69} &
	\multirow{2}{*}{7.01}\\\cline{2-6}
 &
	TR &
	984 &
	666 &
	318 &
	32.3 &
	  &
	  &
	\\\hline
\multirow{2}{*}{AF} &
	SC &
	1040 &
	1039 &
	1$^*$ &
	0.1$^*$ &
	\multirow{2}{*}{0} &
	\multirow{2}{*}{10} &
	\multirow{2}{*}{0.96}\\\cline{2-6}
 &
	TR &
	1040 &
	1039 &
	1 &
	0.1 &
	  &
	  &
	\\\hline
 
\multirow{2}{*}{AP} &
	SC &
	1216 &
	1109 &
	107$^*$ &
	8.8$^*$ &
	\multirow{2}{*}{12} &
	\multirow{2}{*}{74} &
	\multirow{2}{*}{6.09}\\\cline{2-6}
 &
	TR &
	1216 &
	1097 &
	119 &
	9.8 &
	  &
	  &
	\\\hline

\multirow{2}{*}{KP} &
	SC &
	406 &
	332 &
	74$^*$ &
	18.2$^*$ &
	\multirow{2}{*}{1} &
	\multirow{2}{*}{3} &
	\multirow{2}{*}{0.74}\\\cline{2-6}
 &
	TR &
	406 &
	331 &
	75 &
	18.5 &
	  &
	  &
	\\\hline
 
\multirow{2}{*}{ON} &
	SC &
	321 &
	236 &
	85$^*$ &
	26.5$^*$ &
	\multirow{2}{*}{31} &
	\multirow{2}{*}{105} &
	\multirow{2}{*}{32.71}\\\cline{2-6}
 &
	TR &
	321 &
	205 &
	116 &
	36.1 &
	  &
	  &
	\\\hline
 
\multirow{2}{*}{SN} &
	SC &
	804 &
	731 &
	73 &
	9.1 &
	\multirow{2}{*}{5} &
	\multirow{2}{*}{99} &
	\multirow{2}{*}{12.31}\\\cline{2-6}
 &
	TR &
	804 &
	736 &
	68$^*$ &
	8.5$^*$ &
	  &
	  &
	\\\hline

\multirow{2}{*}{AVG} &
	SC &
	795,17 &
	687,33 &
	107,83 &
	15.6 &
	\multirow{2}{*}{-} &
	\multirow{2}{*}{-} &
	\multirow{2}{*}{-}\\\cline{2-6}
 &
	TR &
	795,17 &
	679,00 &
	116,17 &
	17.5 &
	  &
	  &
	\\\hline
\multirow{2}{*}{SD} &
	SC &
	360,30 &
	355,94 &
	103,89 &
	11.8 &
	\multirow{2}{*}{-} &
	\multirow{2}{*}{-} &
	\multirow{2}{*}{-}\\\cline{2-6}
 &
	TR &
	360,30 &
	361,53 &
	107,72 &
	14.2 &
	  &
	  &
	  \\\hline
\end{tabular}
}
\end{table}

As seen in Table~\ref{tab:score}, the number of dead mutants per strategy differs. Column Difference points out the difference in the number of dead mutants between TR (Traditional) and SC (Schemata) for each app. In contrast, the column Divergence points out the number of mutants killed in one strategy but not in the other, i.e., showing a flakiness behavior during test set execution. The most stable program and test set was KP, with one difference in the number of dead mutants and three divergent mutants. In only one program, traditional and mutant schemata strategies got the same results (app AF -- AmazeFileManager) regarding the number of dead mutants, but there are 10 divergences between strategies. We got some variation in the number of killed mutants for all the other apps. The divergences are higher than those of the other apps in the case of ON and SN apps. This must be further investigated, but it is probably related to some library or code structure of such programs, which maximizes flakiness, independent of our effort to avoid it. The symbol $^*$ on the Dead and Mutation Score columns of Table~\ref{tab:score} indicates each app's correct number of dead mutants and mutation score.

Due to this difference, we manually analyzed all these discrepancies and observed that they all occurred due to test flakiness. Besides our effort on ignoring flaky tests, as described in Section~\ref{sec:subjects}, we could observe flakiness problems lately during all test set executions.

Flakiness in software testing is a significant issue that undermines the reliability of test suites and builds systems. While much of the research has focused on unit tests, there is a growing interest in understanding flakiness in the context of system testing, particularly UI-based tests in mobile environments~\cite{Ngo23RTFU}. As pointed out by~\citet{Romano21EAUB}, UI-based system tests, including those in mobile environments, are inherently more complex and resource-intensive than traditional unit tests, leading to a higher incidence of flakiness due to larger input spaces and diverse running conditions. \citet{Memon13ATGA} also corroborates the idea that factors such as the operating system, virtual machine, system load, and initial states contribute to the flakiness of GUI-centric software tests, making system testing appear flakier compared to unit testing. In our research, we faced these problems, and to verify the flakiness effect, we rerun all problematic mutants (both the Traditional and the Schemata ones) to certify that there is no behavior difference between them. Therefore, our analysis concludes that mutation operator implementations according to the mutant schemata and traditional strategies are equivalent.

Additionally, it is crucial to observe that the mutation score determined by the Android instrumented test set is deficient, independently of the app (see Table~\ref{tab:score}). On average, the current instrumented test set detects only 15.6\% of possible faults. Considering the mutation operator set corresponds to a good fault model for Android applications, we need additional effort to improve test set quality at the UI level once the number of remaining faults, represented by the mutants, is very high, more than 84\% of possible faults remained undetected by the current test sets.

\subsection{Answer to RQ2}

Research question two (RQ2) compares both mutation testing strategies regarding resource consumption. To do this, we evaluated the time required to execute the mutants in both strategies: Schemata and Traditional.

We performed this experiment using the same set of apps and mutation operators. We ran all the experiments on a single desktop computer, an HP EliteDesk 800, with a Quad-Core Intel Core i7-4790 64-bit processor, 16GB of RAM, and a 1 TB SSD Samsung 870 QVO SSD, running Linux Mint 21.3. Moreover, we ran the tests in an emulator Pixel 6 Pro, size 6.7", resolution 1440x3120, density 560dpi, running Android release name R, API level 30, ABI x86, target Android 11.0. In the emulator advanced settings, we have Emulated Performance, Graphics: Software, Boot option: Cold boot, Multi-Core: 4, RAM: 2GB, VM Heap: 512, Internal Storage: 8GB, SD Card: 512MB.

Table~\ref{tab:execution-time-saving} presents the running time of each application, considering Mutant Schemata (SC) and Traditional (TR) strategies, all generated mutants, and all Android instrumented test cases available for each app. Execution time is presented in seconds, minutes, hours, and days. The last column of Table~\ref{tab:execution-time-saving} presents the percentage of saving time of the fastest strategy over the slowest one for each app.

The applications that ran the fastest, AF and KP, took around half a day to finish the execution, independently of the strategy, schemata, or traditional used. Still, schemata ran 4.36\% and 4.83 \% faster than traditional for these apps, respectively.

The slowest application was AntennaPod -- AP, which took 5.58 days to execute mutants using schemata and more than 6 days to finish using traditional mutations. Therefore, the schemata strategy saved around 7.56\% of running time for this app.

For all apps, schemata ran faster than traditional, with an average running time of around 6.45\% below the traditional strategy.

\begin{table}[!htb]
    \caption{Time spent running schemata and traditional mutants}
    \label{tab:execution-time-saving}
    \centering
\resizebox{\columnwidth}{!}{%    
    \begin{tabular}{|l|l|r|r|r|r|r|}\hline
\multirow{2}{*}{\bf App} &
	\bf Mutation &
	\multicolumn{5}{|c|}{\bf Total Execution Time}\\\cline{3-7}
 &
	\bf Type &
	\bf Seconds &
	\bf Minutes &
	\bf Hours &
	\bf Days &
	\bf \% Saving\\\hline
\multirow{2}{*}{AE} &
	SC &
	355370 &
	5922.83 &
	98.71 &
	4.11 &
	6.79\\\cline{2-7}
 &
	TR &
	381273 &
	6354.55 &
	105.91 &
	4.41 &
	0.00\\\hline
\multirow{2}{*}{AF} &
	SC &
	44721 &
	745.35 &
	12.42 &
	0.52 &
	4.36\\\cline{2-7}
 &
	TR &
	46762 &
	779.37 &
	12.99 &
	0.54 &
	0.00\\\hline
\multirow{2}{*}{AP} &
	SC &
	481812 &
	8030.20 &
	133.84 &
	5.58 &
	7.56\\\cline{2-7}
 &
	TR &
	521203 &
	8686.72 &
	144.78 &
	6.03 &
	0.00\\\hline

\multirow{2}{*}{KP} &
	SC &
        47128 &
        785.47	&
        13.09	&
        0.55 &
	4.83\\\cline{2-7}
 &
	TR &
        49520	&
        825.33	&
        13.76	& 
        0.57 &
	0.00\\\hline
 
\multirow{2}{*}{ON} &
	SC &
	77659 &
	1294.32 &
	21.57 &
	0.90 &
	5.73\\\cline{2-7}
 &
	TR &
	82381 &
	1373.02 &
	22.88 &
	0.95 &
	0.00\\\hline

\multirow{2}{*}{SN} &
	SC &
	124741 &
	2079.02 &
	34.65 &
	1.44 &
	9.42\\\cline{2-7}
 &
	TR &
	137707 &
	2295.12 &
	38.25 &
	1.59 &
	0.00\\\hline
\multirow{2}{*}{\bf AVG} &
	SC &
        188571.83 &
        3142.86 &
        52.38 &
        2.18 &
        6.45\\\cline{2-7}
 &
	TR &
        203141.00 &
        3385.68	&
        56.43 &
        2.35 &
	0.00\\\hline
 
\multirow{2}{*}{\bf SD} &
	SC &
        184871.24 &
        3081.19	&
        51.35 &
        2.14 &
        1.88\\\cline{2-7}
 &
	TR &
        199906.40 &
        3331.77	&
        55.53 &
        2.31 &
	0.00\\\hline
    \end{tabular}
}
\end{table}

\begin{figure*}
	\centering
	\begin{subfigure}{\columnwidth}
		\includegraphics[width=\linewidth]{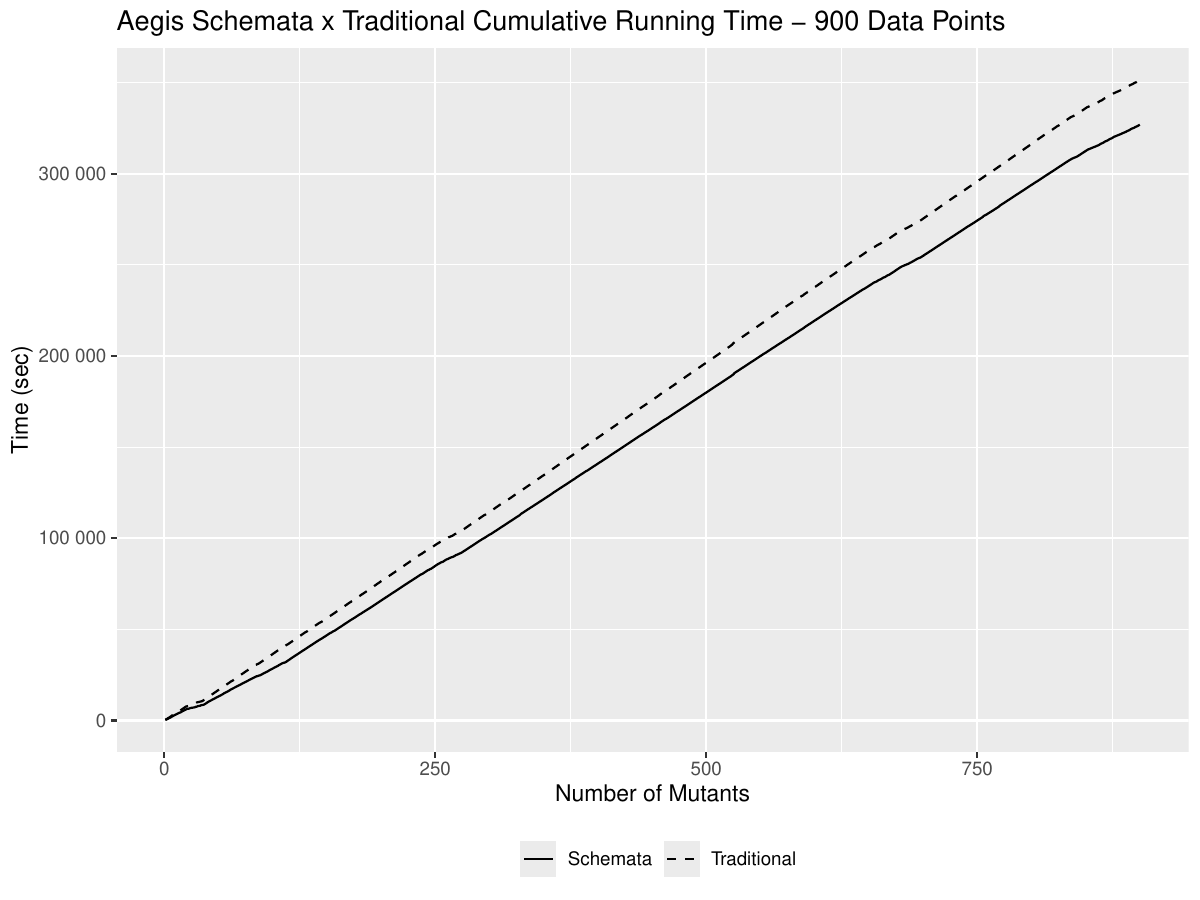}
		\caption{Aegis -- AE}
        \label{fig:crt-aegis}
	\end{subfigure}
	\begin{subfigure}{\columnwidth}
		\includegraphics[width=\linewidth]{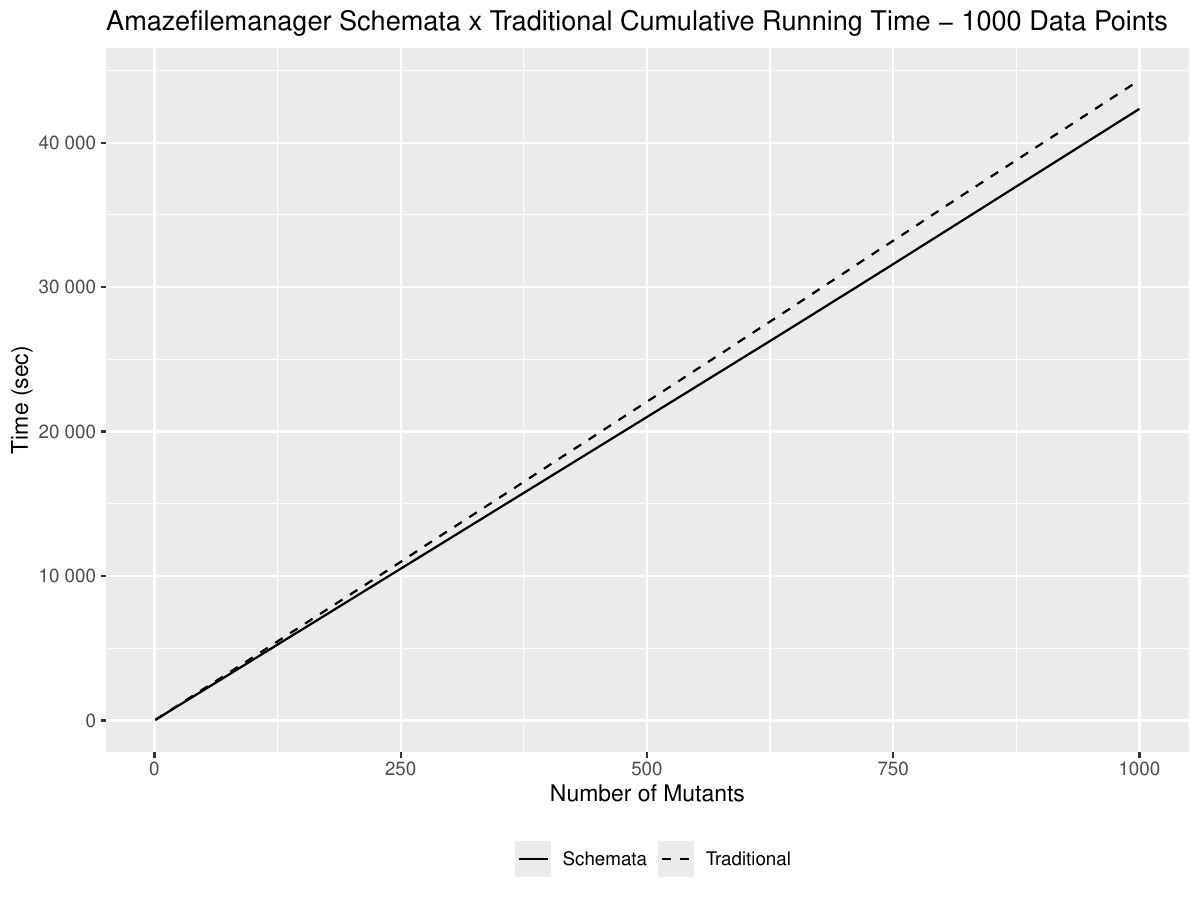}
        \caption{AmazeFileManager -- AM}
        \label{fig:crt-amazefilemanager}
    \end{subfigure}
    \vskip\baselineskip
    \begin{subfigure}{\columnwidth}
        \includegraphics[width=\linewidth]{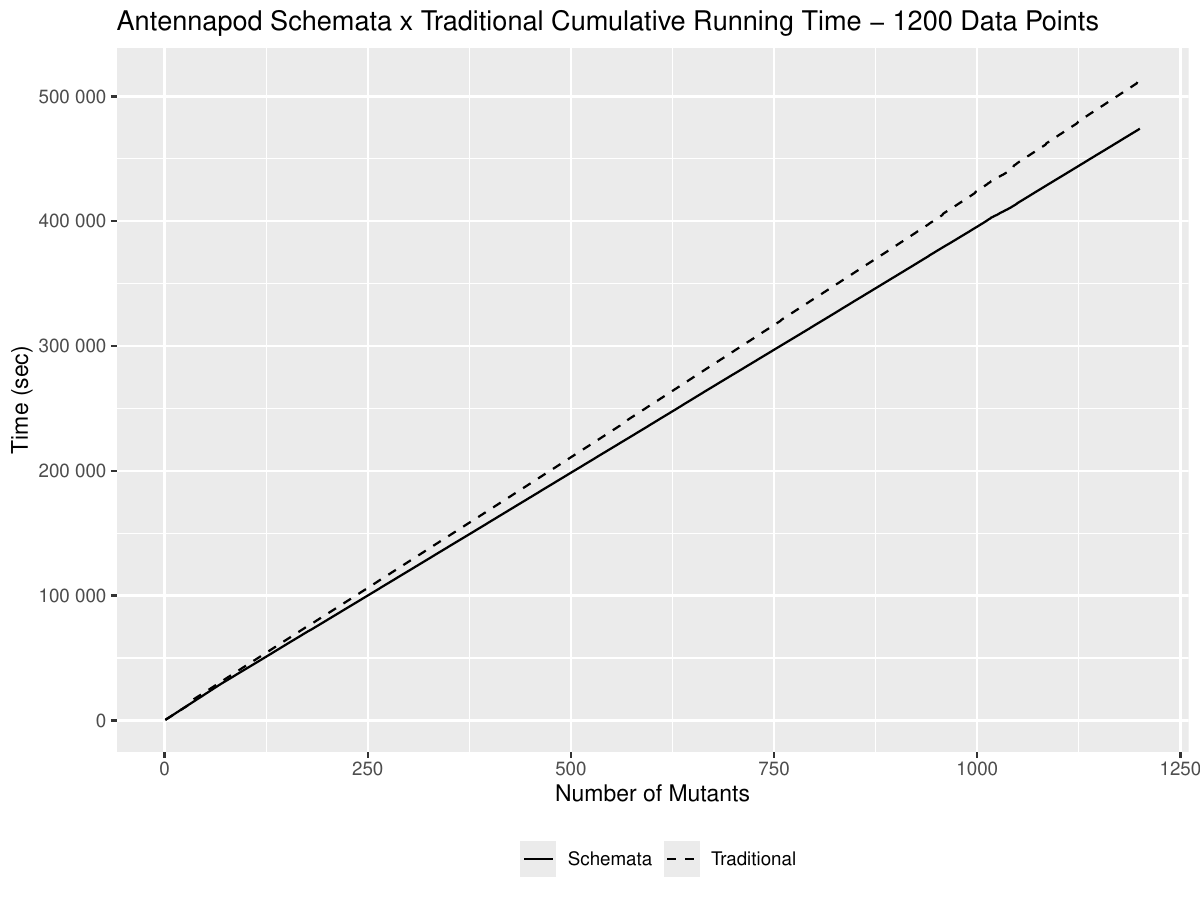}
        \caption{AntennaPod -- AP}
        \label{fig:crt-antennapod}
    \end{subfigure}
    \hfill
    \begin{subfigure}{\columnwidth}
        \includegraphics[width=\linewidth]{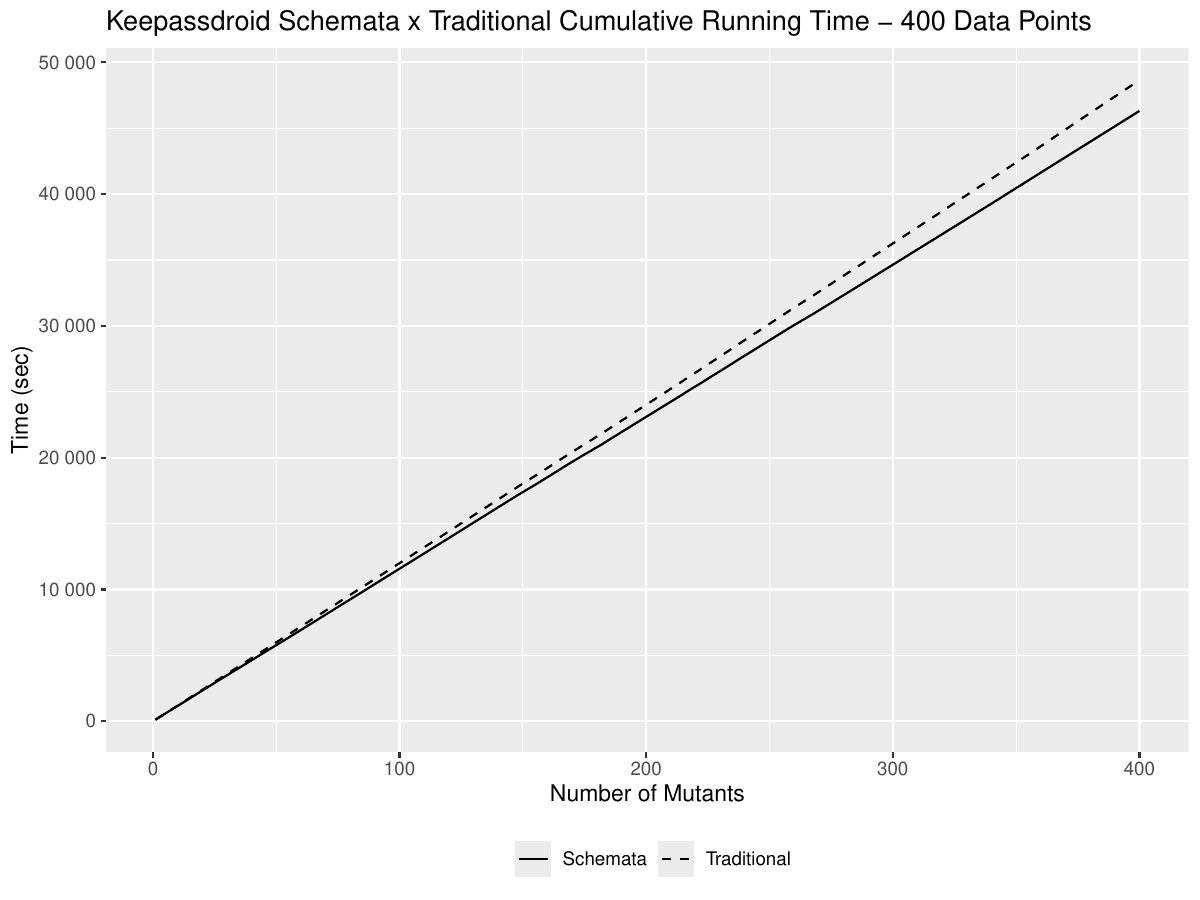}
        \caption{Keepassdroid -- KP}
        \label{fig:crt-keepassdroid}
    \end{subfigure}
    \vskip\baselineskip
    \begin{subfigure}{\columnwidth}
        \includegraphics[width=\linewidth]{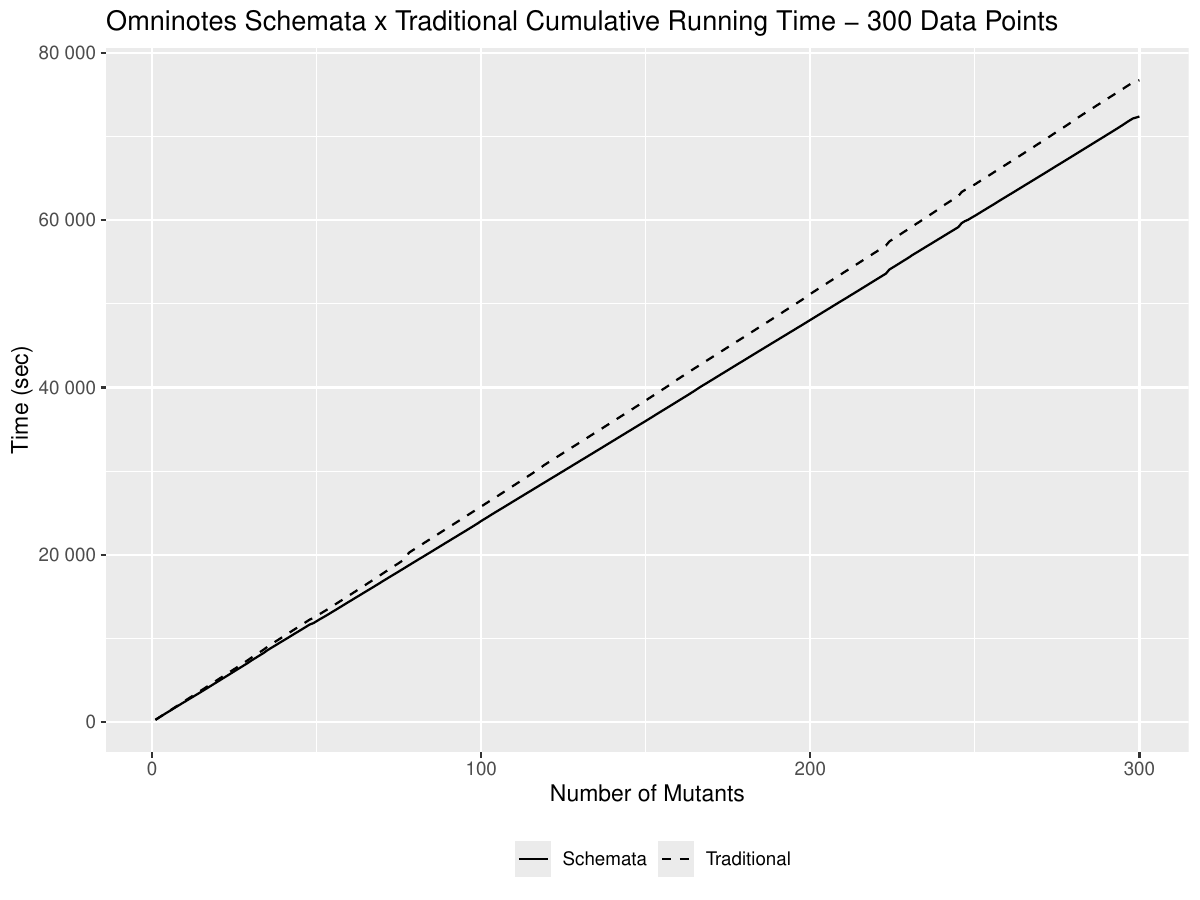}
        \caption{Omni-Notes -- ON}
        \label{fig:crt-omninotes}
    \end{subfigure}
    \hfill
    \begin{subfigure}{\columnwidth}
        \includegraphics[width=\linewidth]{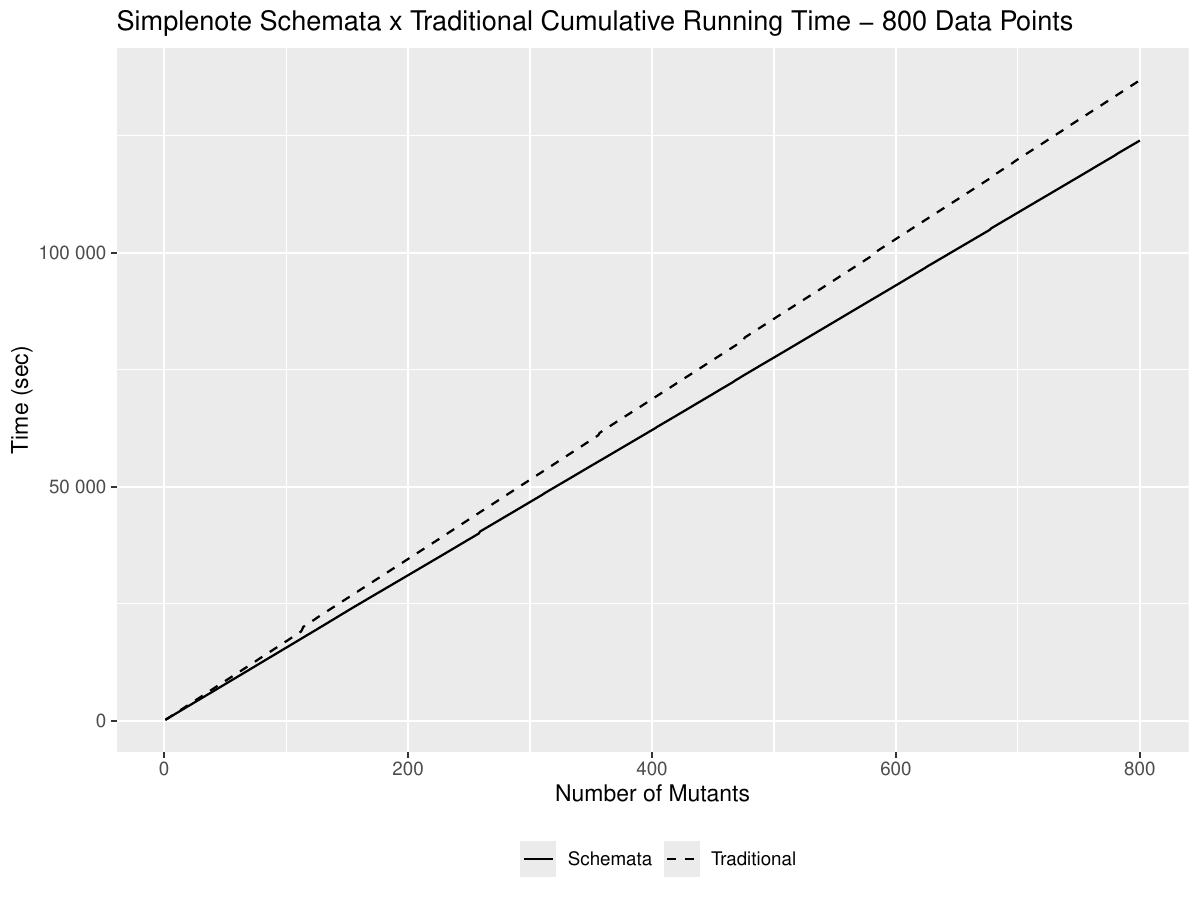}
        \caption{Simplenote -- SN}
        \label{fig:crt-simplenote}
    \end{subfigure}
    \caption{Cumulative running times for various applications}\label{fig:crt}
\end{figure*}

The graphs presented in Figure~\ref{fig:crt} depict the cumulative time of running all mutants against all test cases on Schemata and Traditional mutation strategies. In our experiment, we can observe a linear growth of time, independent of the number of mutants, and the difference in percentage between Schemata and Traditional remains constant over time. %This result contradicts the findings of \citet{https://doi.org/10.1002/stvr.1769}. In their study, the benefits of using Mutant Schema with the All against all approach for a single device are reported to be low if the number of mutants is high. In our study, this gain is constant.

Such a difference in running time results in a time-saving in favor of Schemata between 4.36\% (AF) and 9.42\% (SN), with an average saving of 6.45\% (see Table~\ref{tab:execution-time-saving}).

\begin{figure*}[!htb]
    \centering
    \begin{subfigure}{\columnwidth}
        \includegraphics[width=\linewidth]{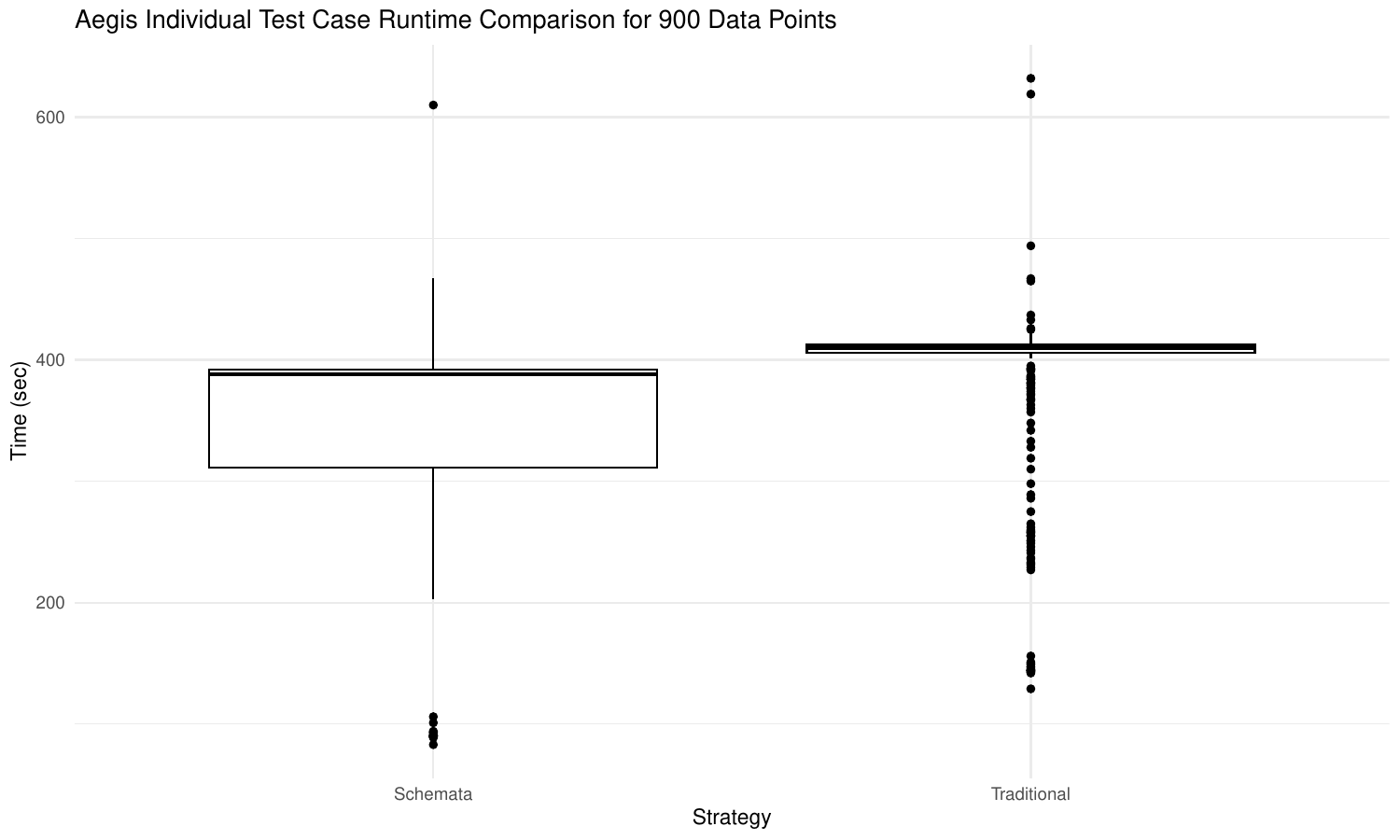}
        \caption{Aegis -- AE}
        \label{fig:tcrt-aegis}
    \end{subfigure}
    \hfill
    \begin{subfigure}{\columnwidth}
        \includegraphics[width=\linewidth]{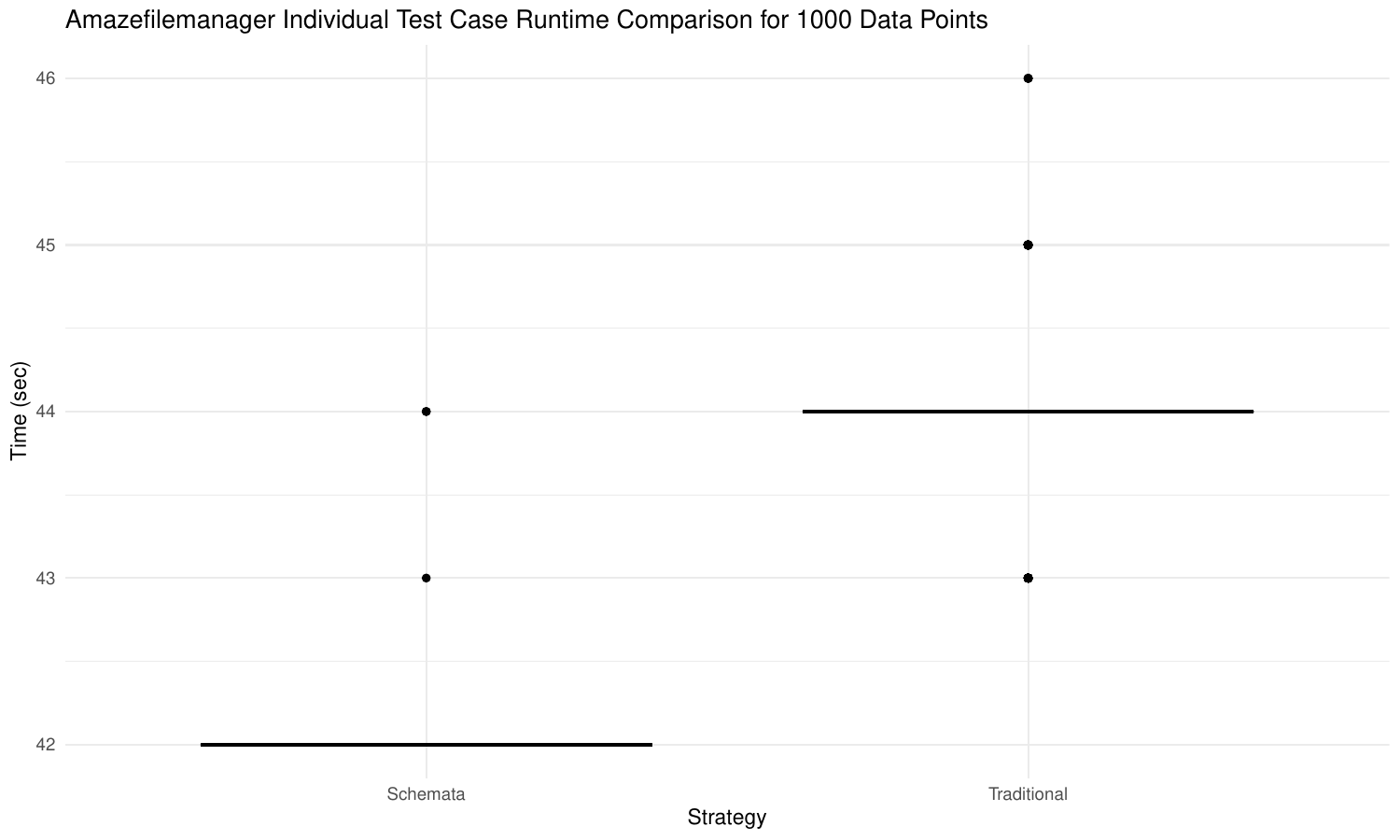}
        \caption{AmazeFileManager -- AM}
        \label{fig:tcrt-amazefilemanager}
    \end{subfigure}
    \vskip\baselineskip
    \begin{subfigure}{\columnwidth}
        \includegraphics[width=\linewidth]{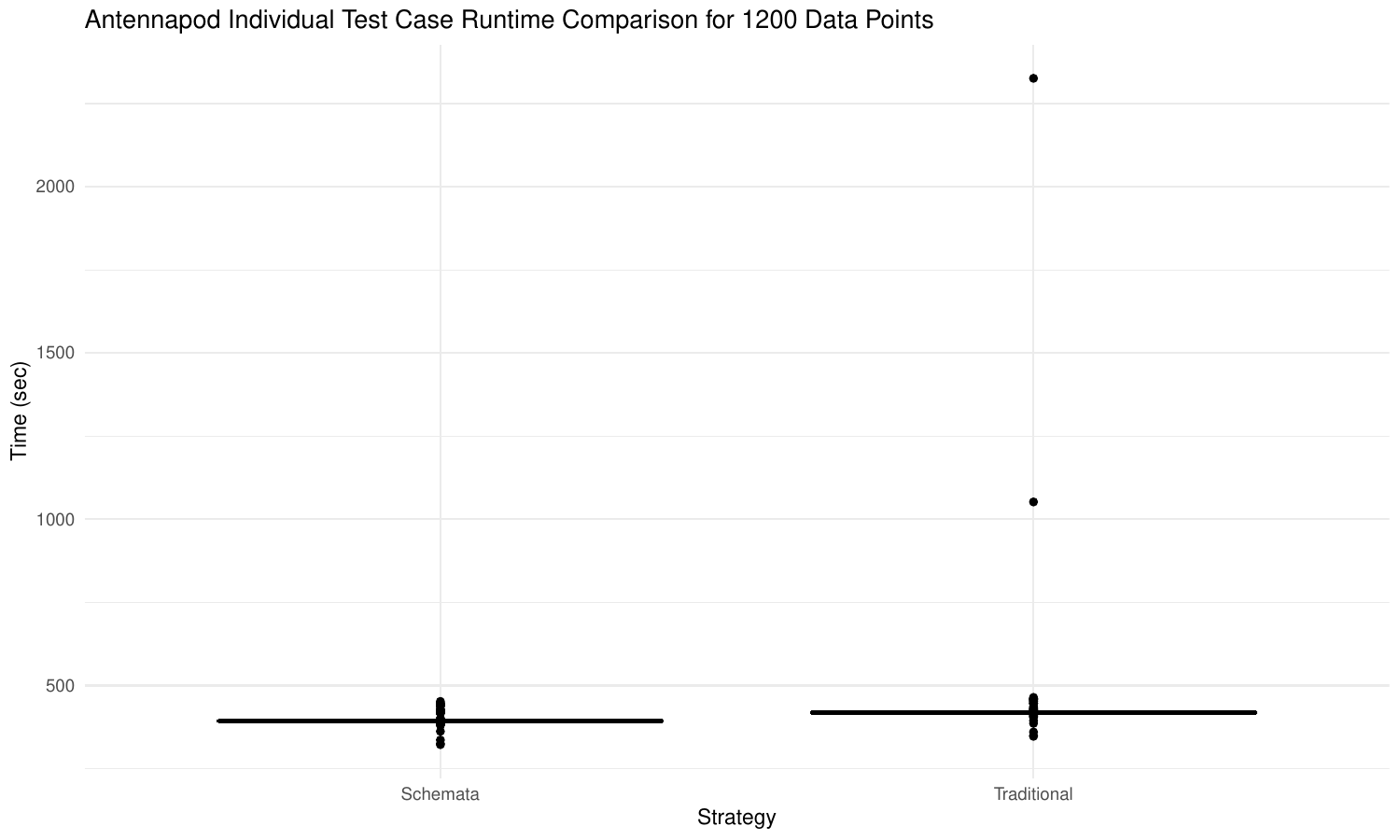}
        \caption{AntennaPod -- AP}
        \label{fig:tcrt-antennapod}
    \end{subfigure}
    \hfill
    \begin{subfigure}{\columnwidth}
        \includegraphics[width=\linewidth]{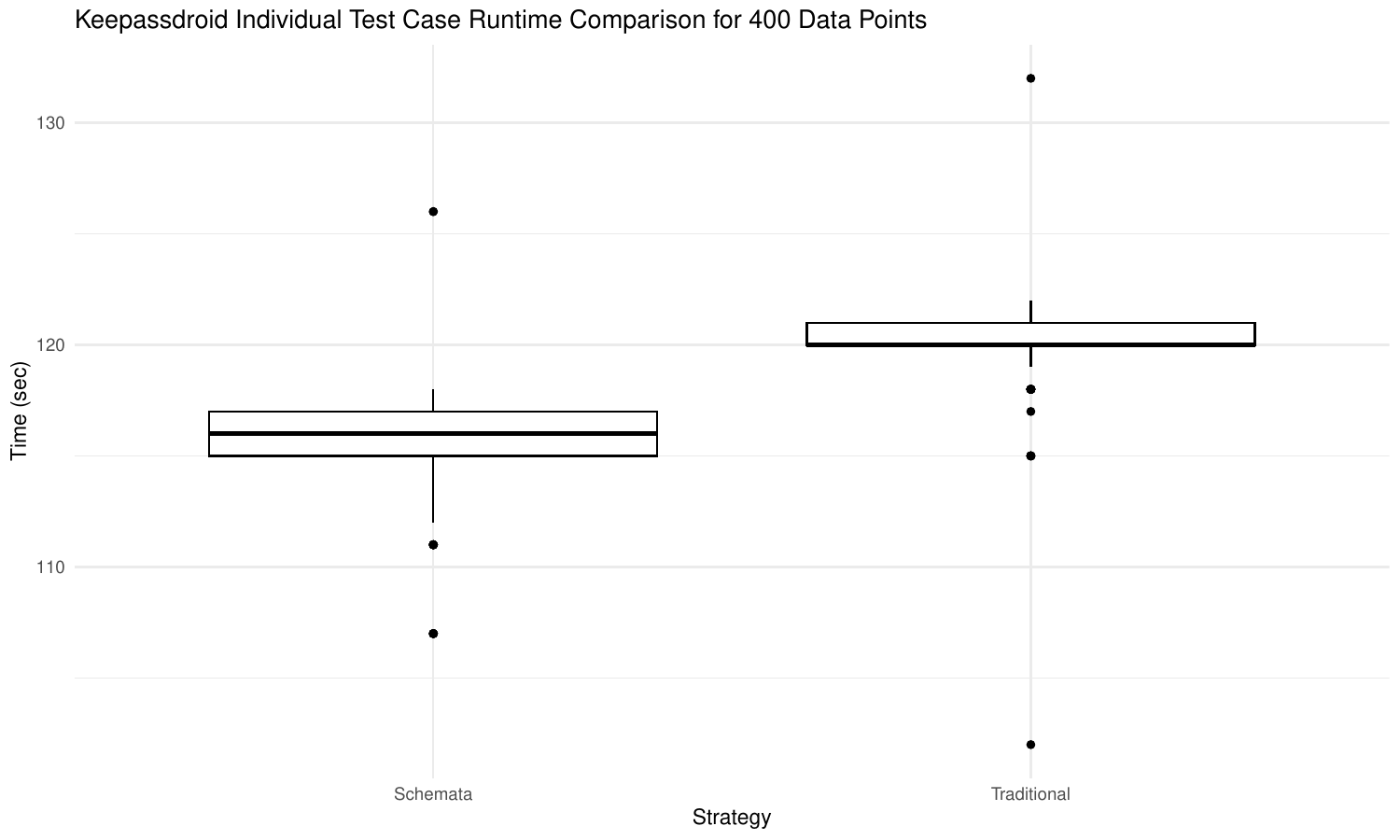}
        \caption{Keepassdroid -- KP}
        \label{fig:tcrt-keepassdroid}
    \end{subfigure}
    \vskip\baselineskip
    \begin{subfigure}{\columnwidth}
        \includegraphics[width=\linewidth]{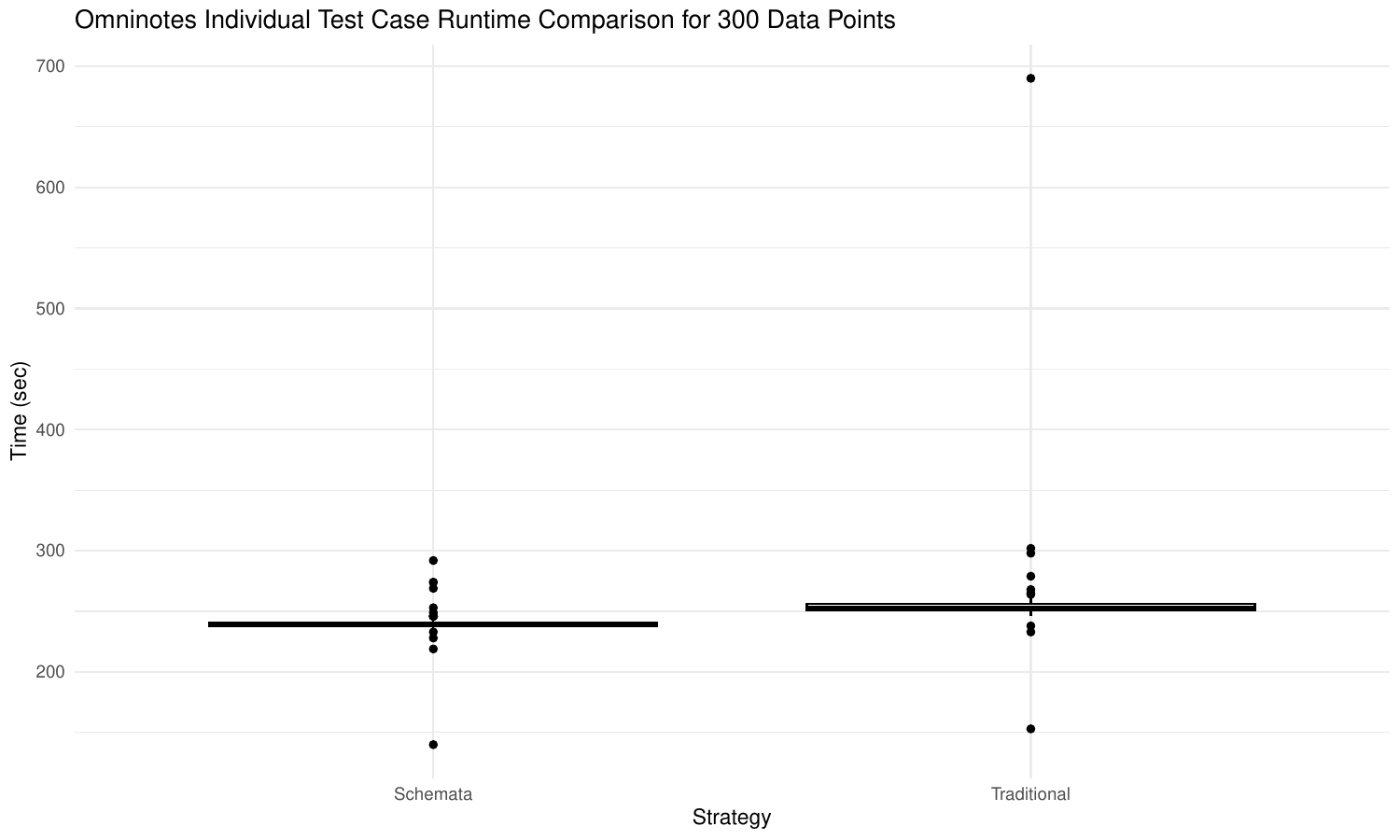}
        \caption{Omni-Notes -- ON}
        \label{fig:tcrt-omninotes}
    \end{subfigure}
    \hfill
    \begin{subfigure}{\columnwidth}
        \includegraphics[width=\linewidth]{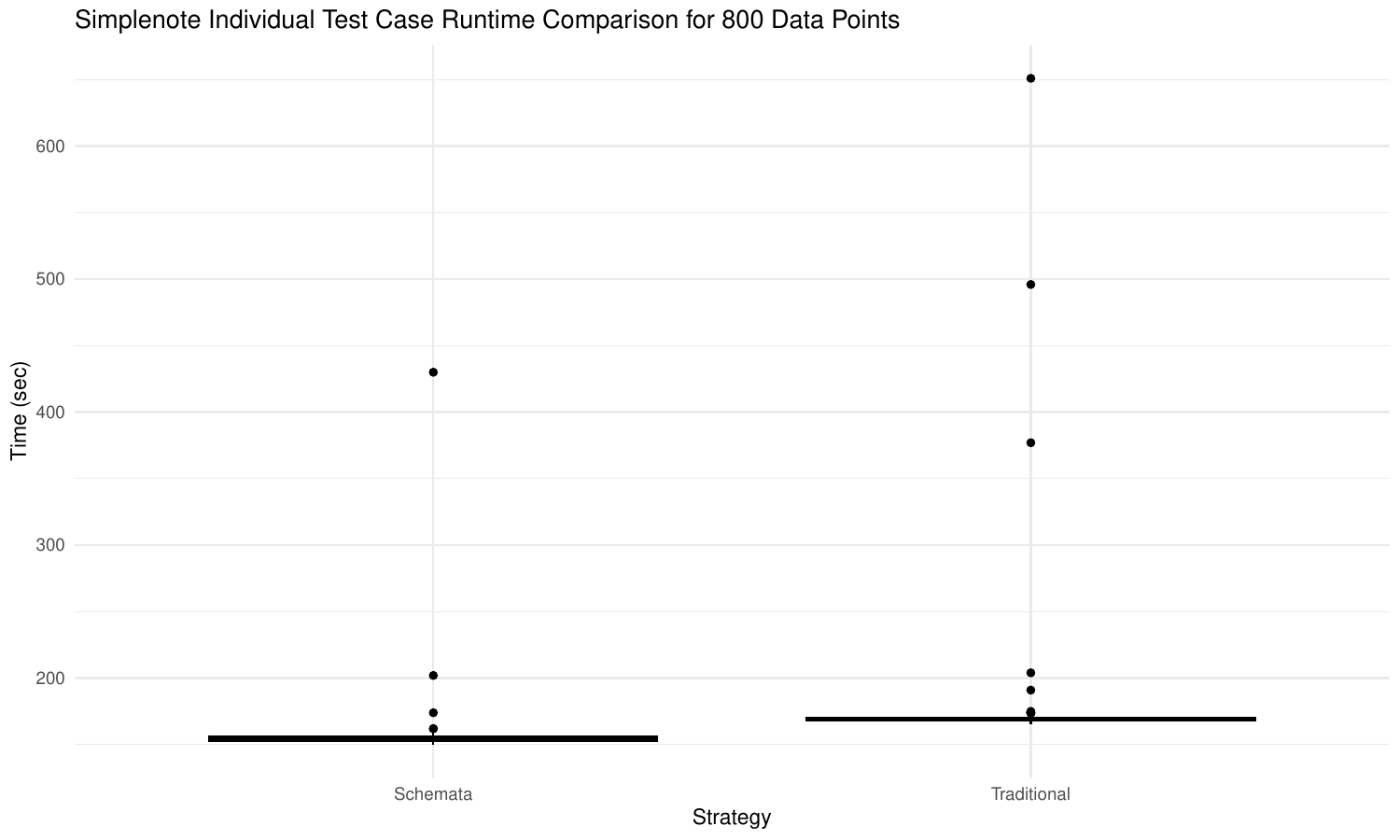}
        \caption{Simplenote -- SN}
        \label{fig:tcrt-simplenote}
    \end{subfigure}
    \caption{Individual test case running times for various applications}\label{fig:tcrt}
\end{figure*}

We also compute the average time for running a single test case on each mutant and plot the graphs in Figure~\ref{fig:tcrt}. Generally, we can also observe that the median of schemata is always below traditional for all apps. The difference is very close in some apps like AP (Figure~\ref{fig:tcrt-antennapod}) and SN (Figure~\ref{fig:tcrt-simplenote}) and larger in others like AM (Figure~\ref{fig:tcrt-amazefilemanager}) and SN (Figure~\ref{fig:tcrt-simplenote}). We can also observe the outliers from these pictures, especially in AE (Figure~\ref{fig:tcrt-aegis}), where it presents more variation in traditional. We believe this is mainly due to flakiness, but this still deserves further investigation, considering the app structure and libraries it uses.

\begin{figure}[!htb]
    \centering
    \includegraphics[width=0.8\columnwidth]{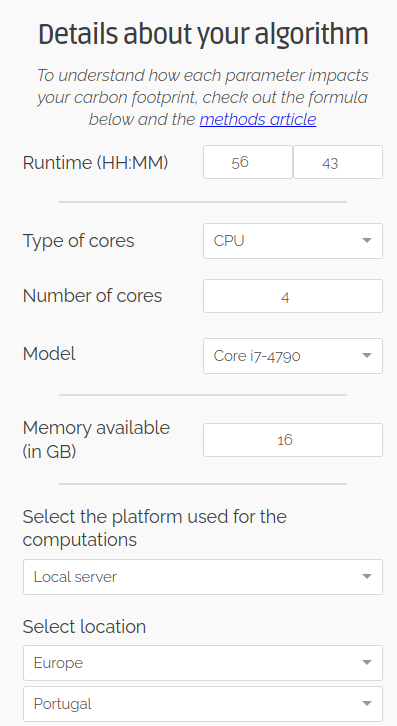}
    \caption{Green Algorithms Calculator parameters.}
    \label{fig:calculator-parameters}
\end{figure}

Finally, to highlight the importance of reducing energy consumption during software development and testing, we consider using the Carbon Footprint calculator\footnote{\url{https://calculator.green-algorithms.org/}}, proposed in the work of~\citet{Lannelongue21GAQC}, to give an idea of the impact of our experiment in terms of carbon footprint and energy consumption. First, we set the calculator parameters considering the machine we have used in our experiment, as depicted in Figure~\ref{fig:calculator-parameters}. 

\begin{figure}[!htb]
    \centering
    \includegraphics[width=\columnwidth]{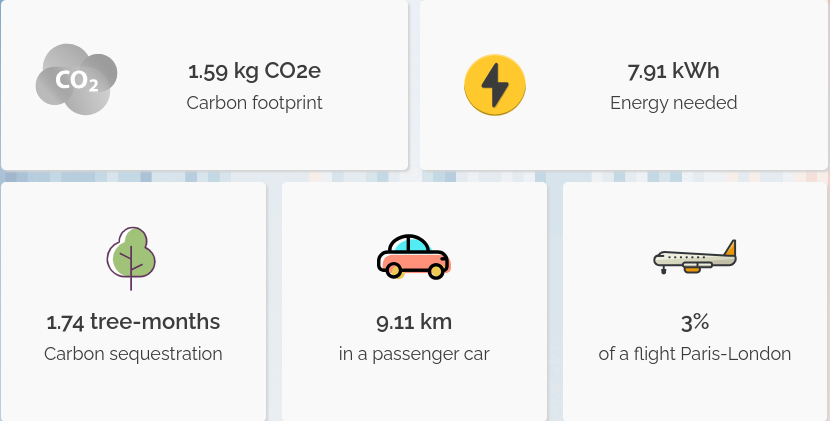}
    \caption{Carbon footprint of mutant schemata strategy.}
    \label{fig:schemata-carbon-footprint}
\end{figure}

\begin{figure}[!htb]
    \centering
    \includegraphics[width=\columnwidth]{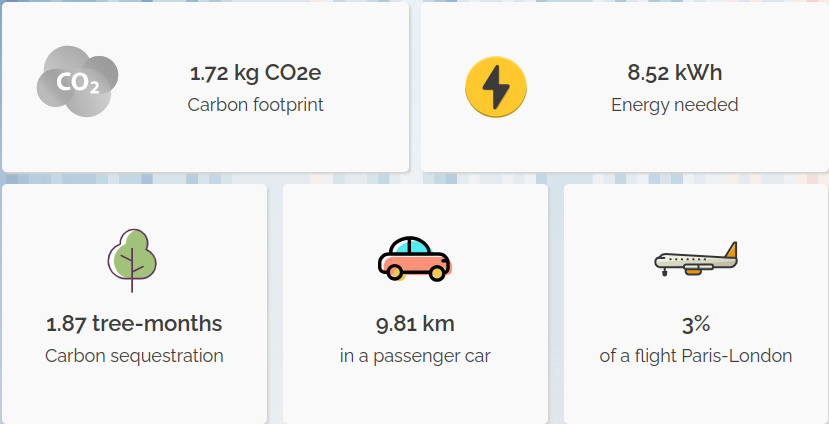}
    \caption{Carbon footprint of traditional mutation strategy.}
    \label{fig:traditional-carbon-footprint}
\end{figure}

Then, we set the average total time spent for running all mutants of Schemata and Traditional mutation strategies, according to the average presented in Table~\ref{tab:execution-time-saving}, and we got the results presented in Figures~\ref{fig:schemata-carbon-footprint}~and~\ref{fig:traditional-carbon-footprint}, corresponding to Schemata and Traditional mutation strategies carbon footprint, respectively.

As observed, the Traditional mutation strategy spent more energy and left a more significant carbon footprint. Table~\ref{tab:green-calculator-data} compares both strategies considering the calculator's five comparison approaches.

\begin{table}[!htb]
    \centering
    \caption{Schemata versus Traditional mutation strategies sustainability comparison.}
    \label{tab:green-calculator-data}
\resizebox{\columnwidth}{!}{%        
    \begin{tabular}{|l|r|r|r|}\hline
Approach &
	Schemata &
	Traditional &
	Difference (\%)\\\hline
Carbon Footprint (CO2e) &
	1.59 &
	1.72 &
	8.18\\\hline
Energy needed (kWh) &
	7.91 &
	8.52 &
	7.71\\\hline
Carbon sequestration (tree/month) &
	1.74 &
	1.87 &
	7.47\\\hline
Car distance (km) &
	9.11 &
	9.81 &
	7.68\\\hline
Flight distance (Paris-London) (\%) &
	3 &
	3 &
	0.00\\\hline
 \end{tabular}
 }
\end{table}

Mutant Schemata is better than Traditional in all other comparison approaches except for flight distance comparison, on which both strategies correspond to the spend of 3\% of fuel between a trip from Paris to London. For instance, the Carbon Footprint of Schemata is 8,18\% lower than Traditional. As defined by~\citet{Lannelongue21GAQC}, the carbon footprint is expressed in Carbon dioxide equivalent (CO2e) and quantifies the global warming potential of a blend of greenhouse gases. It denotes the amount of CO2 that would exert the same global warming effect as the specific mix being considered and serves as a standardized unit for evaluating the environmental impact of human activities. 

Another interesting comparison concerns Carbon sequestration, which is measured in three-months. It corresponds to the amount of CO2 a tree sequesters in one month. This measure indicates how long a mature tree would take to absorb the CO2 emitted by an algorithm~\cite{Lannelongue21GAQC}. In this case, Schemata was 7,47\% better than the Traditional mutation strategy, demanding fewer three-months to neutralize its CO2 emissions.

\begin{figure}[!ht]
    \centering
    \includegraphics[width=\columnwidth]{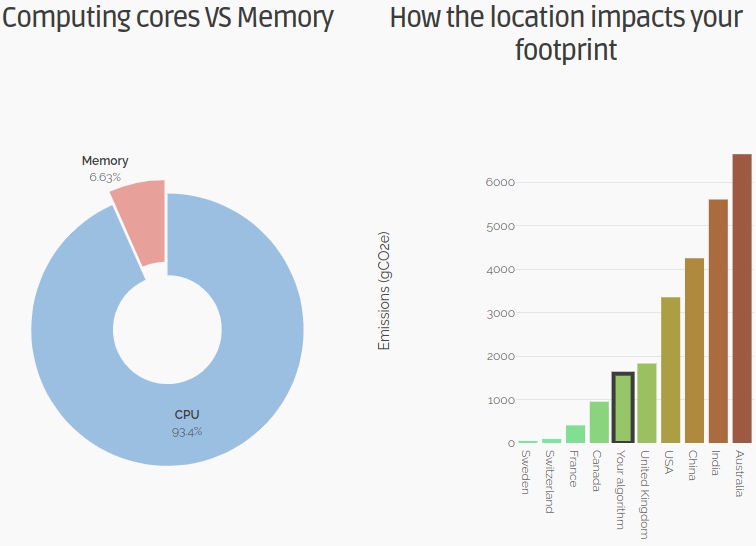}
    \caption{Hardware location impacting carbon footprint.}
    \label{fig:location-experiment-impact}
\end{figure}

Finally, thinking about running experiments in cloud environments or on local servers, the calculator also compares your CPU and memory usage rates considering the computer infrastructure in different regions, as depicted in Figure~\ref{fig:location-experiment-impact}. The difference is related to the energy source in those locations and its corresponding impact on carbon footprint. In our case, as can be observed, Portugal is greening its energy matrix, but there are better options in terms of clean energy, such as Sweden. 

\section{Threats to Validity}\label{sec:threats}

This section discusses factors impacting our research's internal, external, and construct validity.

Regarding construct validity, the execution time may be influenced by the hardware/software characteristics and configuration differences. To mitigate this problem, we used the same machine and ran both mutation operators' strategies (traditional and schemata) on it. Moreover, we fixed compiler and emulator versions on all apps so that they ran on the same machine with the same hardware and software configurations.

In addition, we have implemented scripts to collect the data. Although we may not guarantee the total correctness of these scrips, we tested them over smaller apps, and the tests performed passed.

Regarding internal validity, although we reused the mutation operator's implementations for both traditional and mutant schemata strategies and limited the mutations to only those supported by both approaches, our mutant schemata implementation may not be error-free, so the behavior of the two strategies may not be exactly the same.

However, we mitigated this problem by testing each mutation operator over code snippets and comparing the behavior of some randomly selected mutants (traditional and schemata implementation versions) over real apps.  We analyzed the differences due mainly to flakiness and randomness. This gives confidence that both implement equal behavior.

\revred{Regarding external validity, the results may not be generalizable due to the narrow focus on six apps, so it may be possible to get different results when using other apps.} 
However, we mitigated this problem by using real apps that are publicly available and were chosen based on the listed criteria  \revred{
and have been used in other previous studies.
%The results may not be generalizable regarding external validity due to the narrow focus on six Android apps. However, we mitigated this problem by using real apps that are publicly available and were chosen based on the listed criteria. In addition, the same strategy can be applied to apps of other operating systems/platforms.

Regarding the operators implemented, we} did not implement mutation operators for the entire fault model for Java and Mobile programs discussed and proposed in~\cite{SilvaSTVR22}. However, we covered all 13 Android-specific operators related to GUI and Intent and included one more traditional General-specific arithmetic operator (please refer to Table~\ref{tab:mutation-operators}). Also, it is possible to extend the list of 
mutation operators in the future by implementing the needed scripts and adding them to METFORD.

\revred{
We ran the experiments using \jdk 11, and different results may be obtained from other versions, such as \jdk 17 and \jdk 21. However, we prefer to use \jdk 11 because it demands fewer modifications in the configuration of the app build files; it is the \jdk with the most extensive lifetime, and in our benchmark, we did not detect a significant difference in their performance.

Finally, even though we are using Android apps}, the same \revred{overall} strategy can be applied to apps of other operating systems/platforms. 

\section{Conclusions}\label{sec:conclusions}

This work presented METFORD, a framework supporting mutant schemata and traditional mutation strategies for Android mobile applications. We compared these strategies using METFORD to answer two main research questions.

For RQ1, we verified whether mutants implemented with meta mutant schemata exhibit the same behavior as those with the traditional mutations. Our findings confirmed that both strategies produced consistent mutant behavior. Discrepancies in the number of killed mutants were attributed to test flakiness rather than inherent differences in mutation approaches, indicating the equivalence of the two strategies.

For RQ2, we investigated both implementations' time and disk space resource consumption. The schemata approach consistently demonstrated faster execution times, with savings ranging from 4.36\% to 9.42\%, averaging 6.45\%. This improvement was consistent across all tested applications. Moreover, the schemata method significantly reduced disk space consumption, requiring only a fraction of the space needed for the traditional method and saving more than 99\% of disk space. These findings underscore the efficiency and scalability of the schemata approach, potentially revolutionizing resource management and project scalability in large-scale mutation testing projects.

The implications of our findings are significant for mutation testing in Android applications. The mutant schemata strategy's reduced time and disk space requirements make it a more practical solution for developers, especially for large projects. Its scalability allows for extensive testing without prohibitive costs, making it feasible for real-world applications. However, the low mutation scores observed indicate a need for improved test sets to detect more potential faults, emphasizing the crucial role of developers and testers in future research and improvement. This also suggests the need for better test generation techniques tailored for Android applications, further highlighting the importance of your work in this area.

Our study provides robust evidence supporting the advantages of the mutant schemata strategy, enhancing the efficiency, applicability, and practicality of mutation testing for Android applications. Future work includes:

\begin{itemize}
    \item Optimizing the getMUID() method to reduce overhead;

    \item Improving the Kadabra tool to support different prioritization strategies and enable API execution for DevOps integration;

    \item Developing a LARA compiler for the SMALI language to explore assembly-level mutation testing.
\end{itemize}

Additionally, future research should focus on generating more effective test cases for Android UI fault detection and expanding the experimentation to cover different software and hardware configurations and a broader range of mutation operators.

By addressing these areas, we can significantly enhance the efficiency, applicability, and practicality of mutation testing, making it a more viable option for developers and testers, mainly in the Android ecosystem.

\section*{Acknowledgment}

This work is partially supported by Brazilian Funding Agencies FAPESP (Grant nº 2019/23160-0 and 2023/00001-9), CAPES, and CNPq.

\begin{sloppypar}
\bibliographystyle{elsarticle-num-names}
\bibliography{main}

\begin{thebibliography}{38}
\expandafter\ifx\csname natexlab\endcsname\relax\def\natexlab#1{#1}\fi
\providecommand{\url}[1]{\texttt{#1}}
\providecommand{\href}[2]{#2}
\providecommand{\path}[1]{#1}
\providecommand{\DOIprefix}{doi:}
\providecommand{\ArXivprefix}{arXiv:}
\providecommand{\URLprefix}{URL: }
\providecommand{\Pubmedprefix}{pmid:}
\providecommand{\doi}[1]{\href{http://dx.doi.org/#1}{\path{#1}}}
\providecommand{\Pubmed}[1]{\href{pmid:#1}{\path{#1}}}
\providecommand{\bibinfo}[2]{#2}
\ifx\xfnm\relax \def\xfnm[#1]{\unskip,\space#1}\fi
%Type = Misc
\bibitem[{Buildfire(2023)}]{Buildfire23}
\bibinfo{author}{Buildfire}, \bibinfo{title}{Mobile app download statistics \& usage statistics (2023)}, \bibinfo{howpublished}{Available at url="https://buildfire.com/app-statistics"}, \bibinfo{year}{2023}.
%Type = Article
\bibitem[{Su et~al.(2022)Su, Fan, Chen, Liu, Xu, Pu, and Su}]{9153947}
\bibinfo{author}{T.~Su}, \bibinfo{author}{L.~Fan}, \bibinfo{author}{S.~Chen}, \bibinfo{author}{Y.~Liu}, \bibinfo{author}{L.~Xu}, \bibinfo{author}{G.~Pu}, \bibinfo{author}{Z.~Su},
\newblock \bibinfo{title}{Why my app crashes? understanding and benchmarking framework-specific exceptions of android apps},
\newblock \bibinfo{journal}{IEEE Transactions on Software Engineering} \bibinfo{volume}{48} (\bibinfo{year}{2022}) \bibinfo{pages}{1115--1137}. \DOIprefix\doi{10.1109/TSE.2020.3013438}.
%Type = Article
\bibitem[{Amalfitano et~al.(2022)Amalfitano, Paiva, Inquel, Pinto, Fasolino, and Just}]{10.1145/3526099}
\bibinfo{author}{D.~Amalfitano}, \bibinfo{author}{A.~C.~R. Paiva}, \bibinfo{author}{A.~Inquel}, \bibinfo{author}{L.~Pinto}, \bibinfo{author}{A.~R. Fasolino}, \bibinfo{author}{R.~Just},
\newblock \bibinfo{title}{How do java mutation tools differ?},
\newblock \bibinfo{journal}{Commun. ACM} \bibinfo{volume}{65} (\bibinfo{year}{2022}) \bibinfo{pages}{74–89}. \URLprefix \url{https://doi.org/10.1145/3526099}. \DOIprefix\doi{10.1145/3526099}.
%Type = Inproceedings
\bibitem[{Deng et~al.(2017)Deng, Offutt, and Samudio}]{8009912}
\bibinfo{author}{L.~Deng}, \bibinfo{author}{J.~Offutt}, \bibinfo{author}{D.~Samudio},
\newblock \bibinfo{title}{Is mutation analysis effective at testing android apps?},
\newblock in: \bibinfo{booktitle}{2017 IEEE International Conference on Software Quality, Reliability and Security (QRS)}, \bibinfo{year}{2017}, pp. \bibinfo{pages}{86--93}. \DOIprefix\doi{10.1109/QRS.2017.19}.
%Type = Inproceedings
\bibitem[{Bokaei and Reza~Keyvanpour(2019)}]{Bokaei19}
\bibinfo{author}{N.~N. Bokaei}, \bibinfo{author}{M.~Reza~Keyvanpour},
\newblock \bibinfo{title}{A comparative study of whole issues and challenges in mutation testing},
\newblock in: \bibinfo{booktitle}{2019 5th Conference on Knowledge Based Engineering and Innovation (KBEI)}, \bibinfo{year}{2019}, pp. \bibinfo{pages}{745--754}. \DOIprefix\doi{10.1109/KBEI.2019.8735019}.
%Type = Book
\bibitem[{Wong~(Editor)(2001)}]{Wong01MTNC}
\bibinfo{author}{W.~E. Wong~(Editor)}, \bibinfo{title}{Mutation {Testing} for the {New} {Century}}, Advances in {Database} {Systems}, \bibinfo{publisher}{Springer}, \bibinfo{address}{New York, NY}, \bibinfo{year}{2001}. \URLprefix \url{https://doi.org/10.1007/978-1-4757-5939-6}, \bibinfo{note}{{Springer Science \& Business Media}}.
%Type = Inbook
\bibitem[{Offutt and Untch(2001)}]{Offutt2001}
\bibinfo{author}{A.~J. Offutt}, \bibinfo{author}{R.~H. Untch}, \bibinfo{title}{Mutation 2000: Uniting the Orthogonal}, \bibinfo{publisher}{Springer US}, \bibinfo{address}{Boston, MA}, \bibinfo{year}{2001}, pp. \bibinfo{pages}{34--44}. \URLprefix \url{https://doi.org/10.1007/978-1-4757-5939-6\_7}. \DOIprefix\doi{10.1007/978-1-4757-5939-6\_7}.
%Type = Inproceedings
\bibitem[{Deng and Offutt(2018)}]{DBLP:conf/seke/DengO18}
\bibinfo{author}{L.~Deng}, \bibinfo{author}{J.~Offutt},
\newblock \bibinfo{title}{Reducing the cost of android mutation testing},
\newblock in: \bibinfo{editor}{{\'{O}}.~M. Pereira} (Ed.), \bibinfo{booktitle}{The 30th International Conference on Software Engineering and Knowledge Engineering, Hotel Pullman, Redwood City, California, USA, July 1-3, 2018}, \bibinfo{publisher}{{KSI} Research Inc. and Knowledge Systems Institute Graduate School}, \bibinfo{year}{2018}, pp. \bibinfo{pages}{542--541}. \URLprefix \url{https://doi.org/10.18293/SEKE2018-184}. \DOIprefix\doi{10.18293/SEKE2018-184}.
%Type = Article
\bibitem[{Silva et~al.(2022)Silva, Antonio~do Prado~Lima, Vergilio, and Endo}]{SilvaSTVR22}
\bibinfo{author}{H.~Silva}, \bibinfo{author}{J.~Antonio~do Prado~Lima}, \bibinfo{author}{S.~Vergilio}, \bibinfo{author}{A.~Endo},
\newblock \bibinfo{title}{A mapping study on mutation testing for mobile applications},
\newblock \bibinfo{journal}{Software Testing Verification and Reliability} \bibinfo{volume}{32} (\bibinfo{year}{2022}). \DOIprefix\doi{10.1002/stvr.1801}.
%Type = Inproceedings
\bibitem[{Usaola et~al.(2017)Usaola, Rojas, Rodríguez, and Hernández}]{7899048}
\bibinfo{author}{M.~P. Usaola}, \bibinfo{author}{G.~Rojas}, \bibinfo{author}{I.~Rodríguez}, \bibinfo{author}{S.~Hernández},
\newblock \bibinfo{title}{An architecture for the development of mutation operators},
\newblock in: \bibinfo{booktitle}{2017 IEEE International Conference on Software Testing, Verification and Validation Workshops (ICSTW)}, \bibinfo{year}{2017}, pp. \bibinfo{pages}{143--148}. \DOIprefix\doi{10.1109/ICSTW.2017.31}.
%Type = Inproceedings
\bibitem[{Liu et~al.(2020)Liu, Xiao, Xu, Dou, and Podgurski}]{10.1145/3377812.3382134}
\bibinfo{author}{J.~Liu}, \bibinfo{author}{X.~Xiao}, \bibinfo{author}{L.~Xu}, \bibinfo{author}{L.~Dou}, \bibinfo{author}{A.~Podgurski},
\newblock \bibinfo{title}{Droidmutator: An effective mutation analysis tool for android applications},
\newblock in: \bibinfo{booktitle}{Proceedings of the ACM/IEEE 42nd International Conference on Software Engineering: Companion Proceedings}, ICSE '20, \bibinfo{publisher}{Association for Computing Machinery}, \bibinfo{address}{New York, NY, USA}, \bibinfo{year}{2020}, p. \bibinfo{pages}{77–80}. \URLprefix \url{https://doi.org/10.1145/3377812.3382134}. \DOIprefix\doi{10.1145/3377812.3382134}.
%Type = Inproceedings
\bibitem[{Luna and Ariss(2018)}]{8645883}
\bibinfo{author}{E.~Luna}, \bibinfo{author}{O.~E. Ariss},
\newblock \bibinfo{title}{Edroid: A mutation tool for android apps},
\newblock in: \bibinfo{booktitle}{2018 6th International Conference in Software Engineering Research and Innovation (CONISOFT)}, \bibinfo{year}{2018}, pp. \bibinfo{pages}{99--108}. \DOIprefix\doi{10.1109/CONISOFT.2018.8645883}.
%Type = Inproceedings
\bibitem[{Moran et~al.(2018)Moran, Tufano, Bernal-C\'{a}rdenas, Linares-V\'{a}squez, Bavota, Vendome, Di~Penta, and Poshyvanyk}]{10.1145/3183440.3183492}
\bibinfo{author}{K.~Moran}, \bibinfo{author}{M.~Tufano}, \bibinfo{author}{C.~Bernal-C\'{a}rdenas}, \bibinfo{author}{M.~Linares-V\'{a}squez}, \bibinfo{author}{G.~Bavota}, \bibinfo{author}{C.~Vendome}, \bibinfo{author}{M.~Di~Penta}, \bibinfo{author}{D.~Poshyvanyk},
\newblock \bibinfo{title}{Mdroid+: A mutation testing framework for android},
\newblock in: \bibinfo{booktitle}{Proceedings of the 40th International Conference on Software Engineering: Companion Proceeedings}, ICSE '18, \bibinfo{publisher}{Association for Computing Machinery}, \bibinfo{address}{New York, NY, USA}, \bibinfo{year}{2018}, p. \bibinfo{pages}{33–36}. \URLprefix \url{https://doi.org/10.1145/3183440.3183492}. \DOIprefix\doi{10.1145/3183440.3183492}.
%Type = Inproceedings
\bibitem[{Jabbarvand and Malek(2017)}]{10.1145/3106237.3106244}
\bibinfo{author}{R.~Jabbarvand}, \bibinfo{author}{S.~Malek},
\newblock \bibinfo{title}{µdroid: An energy-aware mutation testing framework for android},
\newblock in: \bibinfo{booktitle}{Proceedings of the 2017 11th Joint Meeting on Foundations of Software Engineering}, ESEC/FSE 2017, \bibinfo{publisher}{Association for Computing Machinery}, \bibinfo{address}{New York, NY, USA}, \bibinfo{year}{2017}, p. \bibinfo{pages}{208–219}. \URLprefix \url{https://doi.org/10.1145/3106237.3106244}. \DOIprefix\doi{10.1145/3106237.3106244}.
%Type = Inproceedings
\bibitem[{Escobar{-}Velasquez et~al.(2019)Escobar{-}Velasquez, Osorio{-}Ria{\~{n}}o, and Linares{-}V{\'{a}}squez}]{DBLP:conf/kbse/Escobar-Velasquez19}
\bibinfo{author}{C.~Escobar{-}Velasquez}, \bibinfo{author}{M.~Osorio{-}Ria{\~{n}}o}, \bibinfo{author}{M.~Linares{-}V{\'{a}}squez},
\newblock \bibinfo{title}{Mutapk: Source-codeless mutant generation for android apps},
\newblock in: \bibinfo{booktitle}{34th {IEEE/ACM} International Conference on Automated Software Engineering, {ASE} 2019, San Diego, CA, USA, November 11-15, 2019}, \bibinfo{publisher}{{IEEE}}, \bibinfo{year}{2019}, pp. \bibinfo{pages}{1090--1093}. \URLprefix \url{https://doi.org/10.1109/ASE.2019.00109}. \DOIprefix\doi{10.1109/ASE.2019.00109}.
%Type = Inproceedings
\bibitem[{Escobar-Vel\'{a}squez et~al.(2020)Escobar-Vel\'{a}squez, Riveros, and Linares-V\'{a}squez}]{EscobarVelasquez20MAPK2ATRMTAA}
\bibinfo{author}{C.~Escobar-Vel\'{a}squez}, \bibinfo{author}{D.~Riveros}, \bibinfo{author}{M.~Linares-V\'{a}squez},
\newblock \bibinfo{title}{Mutapk 2.0: a tool for reducing mutation testing effort of android apps},
\newblock in: \bibinfo{booktitle}{Proceedings of the 28th ACM Joint Meeting on European Software Engineering Conference and Symposium on the Foundations of Software Engineering}, ESEC/FSE 2020, \bibinfo{publisher}{Association for Computing Machinery}, \bibinfo{address}{New York, NY, USA}, \bibinfo{year}{2020}, p. \bibinfo{pages}{1611–1615}.
%Type = Article
\bibitem[{Deng et~al.(2017)Deng, Offutt, Ammann, and Mirzaei}]{Deng15TMAAA}
\bibinfo{author}{L.~Deng}, \bibinfo{author}{J.~Offutt}, \bibinfo{author}{P.~Ammann}, \bibinfo{author}{N.~Mirzaei},
\newblock \bibinfo{title}{Mutation operators for testing android apps},
\newblock \bibinfo{journal}{Information and Software Technology} \bibinfo{volume}{81} (\bibinfo{year}{2017}) \bibinfo{pages}{154--168}. \URLprefix \url{https://www.sciencedirect.com/science/article/pii/S0950584916300684}. \DOIprefix\doi{https://doi.org/10.1016/j.infsof.2016.04.012}.
%Type = Misc
\bibitem[{Wei(2015)}]{Wei15MDMTAA}
\bibinfo{author}{Y.~Wei}, \bibinfo{title}{Mudroid: Mutation testing for android apps}, \bibinfo{year}{2015}. \bibinfo{note}{Available at: \url{http://web4.cs.ucl.ac.uk/staff/Y.Jia/resources/studentprojects/Yuan\_Wei\_MuDroid\_Mutation\_Testing\_for\_Android\_Apps.pdf}}.
%Type = Inproceedings
\bibitem[{Untch et~al.(1993)Untch, Offutt, and Harrold}]{Untch1993MutationAU}
\bibinfo{author}{R.~H. Untch}, \bibinfo{author}{A.~J. Offutt}, \bibinfo{author}{M.~J. Harrold},
\newblock \bibinfo{title}{Mutation analysis using mutant schemata},
\newblock in: \bibinfo{booktitle}{Proceedings of the 1993 ACM SIGSOFT International Symposium on Software Testing and Analysis}, ISSTA '93, \bibinfo{publisher}{Association for Computing Machinery}, \bibinfo{address}{New York, NY, USA}, \bibinfo{year}{1993}, p. \bibinfo{pages}{139–148}. \URLprefix \url{https://doi.org/10.1145/154183.154265}. \DOIprefix\doi{10.1145/154183.154265}.
%Type = Article
\bibitem[{Polo-Usaola and Rodríguez-Trujillo(2021)}]{https://doi.org/10.1002/stvr.1769}
\bibinfo{author}{M.~Polo-Usaola}, \bibinfo{author}{I.~Rodríguez-Trujillo},
\newblock \bibinfo{title}{Analysing the combination of cost reduction techniques in android mutation testing},
\newblock \bibinfo{journal}{Software Testing, Verification and Reliability} \bibinfo{volume}{31} (\bibinfo{year}{2021}) \bibinfo{pages}{e1769}. \URLprefix \url{https://onlinelibrary.wiley.com/doi/abs/10.1002/stvr.1769}. \DOIprefix\doi{https://doi.org/10.1002/stvr.1769}, \bibinfo{note}{e1769 STVR-19-0011.R3}.
%Type = Article
\bibitem[{Pizzoleto et~al.(2019)Pizzoleto, Ferrari, Offutt, Fernandes, and Ribeiro}]{Pizzoleto19ASLRTMRCMT}
\bibinfo{author}{A.~V. Pizzoleto}, \bibinfo{author}{F.~C. Ferrari}, \bibinfo{author}{J.~Offutt}, \bibinfo{author}{L.~Fernandes}, \bibinfo{author}{M.~Ribeiro},
\newblock \bibinfo{title}{A systematic literature review of techniques and metrics to reduce the cost of mutation testing},
\newblock \bibinfo{journal}{Journal of Systems and Software} \bibinfo{volume}{157} (\bibinfo{year}{2019}) \bibinfo{pages}{110388}. \URLprefix \url{https://www.sciencedirect.com/science/article/pii/S0164121219301554}. \DOIprefix\doi{https://doi.org/10.1016/j.jss.2019.07.100}.
%Type = Article
\bibitem[{Escobar-Velásquez et~al.(2022)Escobar-Velásquez, Linares-Vásquez, Bavota, Tufano, Moran, Di~Penta, Vendome, Bernal-Cárdenas, and Poshyvanyk}]{LinarezVasquez17EMTAA}
\bibinfo{author}{C.~Escobar-Velásquez}, \bibinfo{author}{M.~Linares-Vásquez}, \bibinfo{author}{G.~Bavota}, \bibinfo{author}{M.~Tufano}, \bibinfo{author}{K.~Moran}, \bibinfo{author}{M.~Di~Penta}, \bibinfo{author}{C.~Vendome}, \bibinfo{author}{C.~Bernal-Cárdenas}, \bibinfo{author}{D.~Poshyvanyk},
\newblock \bibinfo{title}{Enabling mutant generation for open- and closed-source android apps},
\newblock \bibinfo{journal}{IEEE Transactions on Software Engineering} \bibinfo{volume}{48} (\bibinfo{year}{2022}) \bibinfo{pages}{186--208}. \DOIprefix\doi{10.1109/TSE.2020.2982638}.
%Type = Inproceedings
\bibitem[{Kurtz et~al.(2016)Kurtz, Ammann, Offutt, Delamaro, Kurtz, and G\"{o}k\c{c}e}]{Kurtz16AVSM}
\bibinfo{author}{B.~Kurtz}, \bibinfo{author}{P.~Ammann}, \bibinfo{author}{J.~Offutt}, \bibinfo{author}{M.~E. Delamaro}, \bibinfo{author}{M.~Kurtz}, \bibinfo{author}{N.~G\"{o}k\c{c}e},
\newblock \bibinfo{title}{Analyzing the validity of selective mutation with dominator mutants},
\newblock in: \bibinfo{booktitle}{Proceedings of the 2016 24th ACM SIGSOFT International Symposium on Foundations of Software Engineering}, FSE 2016, \bibinfo{publisher}{Association for Computing Machinery}, \bibinfo{address}{New York, NY, USA}, \bibinfo{year}{2016}, p. \bibinfo{pages}{571–582}. \URLprefix \url{https://doi.org/10.1145/2950290.2950322}. \DOIprefix\doi{10.1145/2950290.2950322}.
%Type = Article
\bibitem[{Cardoso et~al.(2016)Cardoso, Coutinho, Carvalho, Diniz, Petrov, Luk, and Gon{\c{c}}alves}]{cardoso2016performance}
\bibinfo{author}{J.~M.~P. Cardoso}, \bibinfo{author}{J.~G. Coutinho}, \bibinfo{author}{T.~Carvalho}, \bibinfo{author}{P.~C. Diniz}, \bibinfo{author}{Z.~Petrov}, \bibinfo{author}{W.~Luk}, \bibinfo{author}{F.~Gon{\c{c}}alves},
\newblock \bibinfo{title}{Performance-driven instrumentation and mapping strategies using the lara aspect-oriented programming approach},
\newblock \bibinfo{journal}{Software: Practice and Experience} \bibinfo{volume}{46} (\bibinfo{year}{2016}) \bibinfo{pages}{251--287}.
%Type = Article
\bibitem[{Pinto et~al.(2018)Pinto, Carvalho, Bispo, Ramalho, and Cardoso}]{pinto2018aspect}
\bibinfo{author}{P.~Pinto}, \bibinfo{author}{T.~Carvalho}, \bibinfo{author}{J.~Bispo}, \bibinfo{author}{M.~A. Ramalho}, \bibinfo{author}{J.~M.~P. Cardoso},
\newblock \bibinfo{title}{Aspect composition for multiple target languages using lara},
\newblock \bibinfo{journal}{Computer Languages, Systems \& Structures} \bibinfo{volume}{53} (\bibinfo{year}{2018}) \bibinfo{pages}{1--26}.
%Type = Article
\bibitem[{Carvalho et~al.(2023)Carvalho, Bispo, Pinto, and Cardoso}]{carvalho2023dsl}
\bibinfo{author}{T.~Carvalho}, \bibinfo{author}{J.~Bispo}, \bibinfo{author}{P.~Pinto}, \bibinfo{author}{J.~M. Cardoso},
\newblock \bibinfo{title}{A dsl-based runtime adaptivity framework for java},
\newblock \bibinfo{journal}{SoftwareX} \bibinfo{volume}{23} (\bibinfo{year}{2023}) \bibinfo{pages}{101496}.
%Type = Article
\bibitem[{Bispo and Cardoso(2020)}]{bispo2020clava}
\bibinfo{author}{J.~Bispo}, \bibinfo{author}{J.~M. Cardoso},
\newblock \bibinfo{title}{Clava: C/c++ source-to-source compilation using lara},
\newblock \bibinfo{journal}{SoftwareX} \bibinfo{volume}{12} (\bibinfo{year}{2020}) \bibinfo{pages}{100565}.
%Type = Article
\bibitem[{Pawlak et~al.(2016)Pawlak, Monperrus, Petitprez, Noguera, and Seinturier}]{https://doi.org/10.1002/spe.2346}
\bibinfo{author}{R.~Pawlak}, \bibinfo{author}{M.~Monperrus}, \bibinfo{author}{N.~Petitprez}, \bibinfo{author}{C.~Noguera}, \bibinfo{author}{L.~Seinturier},
\newblock \bibinfo{title}{Spoon: A library for implementing analyses and transformations of java source code},
\newblock \bibinfo{journal}{Software: Practice and Experience} \bibinfo{volume}{46} (\bibinfo{year}{2016}) \bibinfo{pages}{1155--1179}. \URLprefix \url{https://onlinelibrary.wiley.com/doi/abs/10.1002/spe.2346}. \DOIprefix\doi{https://doi.org/10.1002/spe.2346}.
%Type = Article
\bibitem[{Petrović et~al.(2022)Petrović, Ivanković, Fraser, and Just}]{Petrovic22PMTS}
\bibinfo{author}{G.~Petrović}, \bibinfo{author}{M.~Ivanković}, \bibinfo{author}{G.~Fraser}, \bibinfo{author}{R.~Just},
\newblock \bibinfo{title}{Practical {Mutation} {Testing} at {Scale}: {A} view from {Google}},
\newblock \bibinfo{journal}{IEEE Transactions on Software Engineering} \bibinfo{volume}{48} (\bibinfo{year}{2022}) \bibinfo{pages}{3900--3912}. \DOIprefix\doi{10.1109/TSE.2021.3107634}.
%Type = Inproceedings
\bibitem[{Delamaro et~al.(2018)Delamaro, Chaim, and Maldonado}]{Delamaro18WMMU}
\bibinfo{author}{M.~Delamaro}, \bibinfo{author}{M.~L. Chaim}, \bibinfo{author}{J.~C. Maldonado},
\newblock \bibinfo{title}{Where {Are} the {Minimal} {Mutants}?},
\newblock in: \bibinfo{booktitle}{Proceedings of the {XXXII} {Brazilian} {Symposium} on {Software} {Engineering}}, volume~\bibinfo{volume}{1} of \textit{\bibinfo{series}{{SBES}'18}}, \bibinfo{publisher}{ACM}, \bibinfo{year}{2018}, pp. \bibinfo{pages}{190--195}. \DOIprefix\doi{10.1145/3266237.3266241}, \bibinfo{note}{bibtex*[numpages=6;acmid=3266241] event-place: Sao Carlos, Brazil}.
%Type = Inproceedings
\bibitem[{Dong et~al.(2021)Dong, Tiwari, Yu, and Roychoudhury}]{Dong21FTDAVEOE}
\bibinfo{author}{Z.~Dong}, \bibinfo{author}{A.~Tiwari}, \bibinfo{author}{X.~L. Yu}, \bibinfo{author}{A.~Roychoudhury},
\newblock \bibinfo{title}{Flaky test detection in android via event order exploration},
\newblock in: \bibinfo{booktitle}{Proceedings of the 29th ACM Joint Meeting on European Software Engineering Conference and Symposium on the Foundations of Software Engineering}, ESEC/FSE 2021, \bibinfo{publisher}{Association for Computing Machinery}, \bibinfo{address}{New York, NY, USA}, \bibinfo{year}{2021}, p. \bibinfo{pages}{367–378}.
%Type = Inproceedings
\bibitem[{Xiong et~al.(2023)Xiong, Xu, Su, Sun, Wang, Wen, Pu, He, and Su}]{Xiong23AESFBAA}
\bibinfo{author}{Y.~Xiong}, \bibinfo{author}{M.~Xu}, \bibinfo{author}{T.~Su}, \bibinfo{author}{J.~Sun}, \bibinfo{author}{J.~Wang}, \bibinfo{author}{H.~Wen}, \bibinfo{author}{G.~Pu}, \bibinfo{author}{J.~He}, \bibinfo{author}{Z.~Su},
\newblock \bibinfo{title}{An empirical study of functional bugs in android apps},
\newblock in: \bibinfo{booktitle}{Proceedings of the 32nd ACM SIGSOFT International Symposium on Software Testing and Analysis}, ISSTA 2023, \bibinfo{publisher}{Association for Computing Machinery}, \bibinfo{address}{New York, NY, USA}, \bibinfo{year}{2023}, p. \bibinfo{pages}{1319–1331}.
%Type = Inproceedings
\bibitem[{Pan et~al.(2020)Pan, Huang, Wang, Zhang, and Li}]{Pan20RLBCDTAA}
\bibinfo{author}{M.~Pan}, \bibinfo{author}{A.~Huang}, \bibinfo{author}{G.~Wang}, \bibinfo{author}{T.~Zhang}, \bibinfo{author}{X.~Li},
\newblock \bibinfo{title}{Reinforcement learning based curiosity-driven testing of android applications},
\newblock in: \bibinfo{booktitle}{Proceedings of the 29th ACM SIGSOFT International Symposium on Software Testing and Analysis}, ISSTA 2020, \bibinfo{publisher}{Association for Computing Machinery}, \bibinfo{address}{New York, NY, USA}, \bibinfo{year}{2020}, p. \bibinfo{pages}{153–164}.
%Type = Inproceedings
\bibitem[{Su et~al.(2021)Su, Wang, and Su}]{Su21BAGTAARWB}
\bibinfo{author}{T.~Su}, \bibinfo{author}{J.~Wang}, \bibinfo{author}{Z.~Su},
\newblock \bibinfo{title}{Benchmarking automated gui testing for android against real-world bugs},
\newblock in: \bibinfo{booktitle}{Proceedings of the 29th ACM Joint Meeting on European Software Engineering Conference and Symposium on the Foundations of Software Engineering}, ESEC/FSE 2021, \bibinfo{publisher}{Association for Computing Machinery}, \bibinfo{address}{New York, NY, USA}, \bibinfo{year}{2021}, p. \bibinfo{pages}{119–130}.
%Type = Inproceedings
\bibitem[{Ngo et~al.(2023)Ngo, Nguyen, and Nguyen}]{Ngo23RTFU}
\bibinfo{author}{K.~Ngo}, \bibinfo{author}{V.~Nguyen}, \bibinfo{author}{T.~Nguyen},
\newblock \bibinfo{title}{Research on test flakiness: from unit to system testing},
\newblock in: \bibinfo{booktitle}{Proceedings of the 37th IEEE/ACM International Conference on Automated Software Engineering}, ASE '22, \bibinfo{publisher}{Association for Computing Machinery}, \bibinfo{address}{New York, NY, USA}, \bibinfo{year}{2023}, pp. \bibinfo{pages}{1--4}. \URLprefix \url{https://doi.org/10.1145/3551349.3563242}. \DOIprefix\doi{10.1145/3551349.3563242}.
%Type = Inproceedings
\bibitem[{Romano et~al.(2021)Romano, Song, Grandhi, Yang, and Wang}]{Romano21EAUB}
\bibinfo{author}{A.~Romano}, \bibinfo{author}{Z.~Song}, \bibinfo{author}{S.~Grandhi}, \bibinfo{author}{W.~Yang}, \bibinfo{author}{W.~Wang},
\newblock \bibinfo{title}{An empirical analysis of ui-based flaky tests},
\newblock in: \bibinfo{booktitle}{2021 IEEE/ACM 43rd International Conference on Software Engineering (ICSE)}, \bibinfo{year}{2021}, pp. \bibinfo{pages}{1585--1597}. \DOIprefix\doi{10.1109/ICSE43902.2021.00141}.
%Type = Inproceedings
\bibitem[{Memon and Cohen(2013)}]{Memon13ATGA}
\bibinfo{author}{A.~M. Memon}, \bibinfo{author}{M.~B. Cohen},
\newblock \bibinfo{title}{Automated testing of gui applications: Models, tools, and controlling flakiness},
\newblock in: \bibinfo{booktitle}{2013 35th International Conference on Software Engineering (ICSE)}, \bibinfo{year}{2013}, pp. \bibinfo{pages}{1479--1480}. \DOIprefix\doi{10.1109/ICSE.2013.6606750}.
%Type = Article
\bibitem[{Lannelongue et~al.(2021)Lannelongue, Grealey, and Inouye}]{Lannelongue21GAQC}
\bibinfo{author}{L.~Lannelongue}, \bibinfo{author}{J.~Grealey}, \bibinfo{author}{M.~Inouye},
\newblock \bibinfo{title}{Green {Algorithms}: {Quantifying} the {Carbon} {Footprint} of {Computation}},
\newblock \bibinfo{journal}{Advanced Science} \bibinfo{volume}{8} (\bibinfo{year}{2021}) \bibinfo{pages}{2100707}. \DOIprefix\doi{10.1002/advs.202100707}.

\end{thebibliography}
\end{sloppypar}

\end{document}